\documentclass[fleqn,usenatbib]{mnras}

\usepackage{newtxtext,newtxmath}

\usepackage[T1]{fontenc}

\DeclareRobustCommand{\VAN}[3]{#2}
\let\VANthebibliography\thebibliography
\def\thebibliography{\DeclareRobustCommand{\VAN}[3]{##3}\VANthebibliography}

\usepackage{graphicx}	
\usepackage{amsmath}	

\newcommand{\ks}[1]{\textcolor{black}{#1}}

\title[ICs for WD planetary systems]{
Birth cluster simulations of planetary systems with multiple super-Earths: initial conditions for white dwarf pollution drivers}

\author[K. Stock et al.]{
Katja Stock,$^{1}$\thanks{E-mail: katja.stock@uni-heidelberg.de}\thanks{Fellow of the International Max Planck Research School for Astronomy and Cosmic Physics at the University of Heidelberg (IMPRS-HD).}
Dimitri Veras,$^{2,3,4}$\thanks{STFC Ernest Rutherford Fellow.}
Maxwell X. Cai,$^{5}$
Rainer Spurzem$^{1,6,7}$\thanks{Research Fellow of Frankfurt Institute for Advanced Studies.}
and Simon Portegies Zwart$^{5}$
\\
$^{1}$Astronomisches Rechen-Institut, Zentrum f\"ur Astronomie der Universit\"at Heidelberg, M\"onchhofstra\ss{}e 12-14, D-69120 Heidelberg, Germany\\
$^{2}$Centre for Exoplanets and Habitability, University of Warwick, Coventry CV4 7AL, UK\\
$^{3}$Centre for Space Domain Awareness, University of Warwick, Coventry CV4 7AL, UK\\
$^{4}$Department of Physics, University of Warwick, Coventry CV4 7AL, UK\\
$^{5}$Leiden Observatory, Leiden University, PO Box 9513, NL-2300 RA Leiden, the Netherlands\\
$^{6}$National Astronomical Observatories and Key Laboratory of Computational Astrophysics, Chinese Academy of Sciences, 20A Datun Road, \\Chaoyang District, Beijing 100101, P.R. China\\
$^{7}$Kavli Institute for Astronomy and Astrophysics, Peking University, 5 Yi He Yuan Road, Haidian District, Beijing 100871, P.R. China
}

\date{Accepted XXX. Received YYY; in original form ZZZ}

\pubyear{2015}

\begin{document}
\label{firstpage}
\pagerange{\pageref{firstpage}--\pageref{lastpage}}
\maketitle

\begin{abstract}
Previous investigations have revealed that eccentric super-Earths represent a class of planets which are particularly effective at transporting minor bodies towards white dwarfs and subsequently polluting their atmospheres with observable chemical signatures. However, the lack of discoveries of these planets beyond a few astronomical units from their host stars prompts a better understanding of their orbital architectures from their nascent birth cluster. Here, we perform stellar cluster simulations of 3-planet and 7-planet systems containing super-Earths on initially circular, coplanar orbits. We adopt the typical stellar masses of main-sequence progenitors of white dwarfs ($1.5\,\mathrm{M}_{\odot}$ – $2.5\,\mathrm{M}_{\odot}$) as host stars and include 8,000 main-sequence stars following a Kroupa initial mass function in our clusters. Our results reveal that about 30 per cent of the simulated planets generate eccentricities of at least 0.1 by the time of cluster dissolution, which would aid white dwarf pollution. We provide our output parameters to the community for potential use as initial conditions for subsequent evolution simulations.

\end{abstract}




\section{Introduction}

Over 99 per cent of all known exoplanet host stars will eventually evolve into white dwarfs (WDs). This fact emphasises the importance of being able to connect planetary architectures around WDs to their previous incarnations around giant branch and main-sequence stars, and to the processes which occurred in their nebular birth clusters.

One way to pursue this connection is to consider the observations of known planetary systems around WDs. \cite{Veras2021} partitions these observations into four classes: (i) major planets, (ii) minor planets (such as asteroids, comets or moons, but also the remnants of larger planets), (iii) discs and rings, and (iv) chemical pollution by metals in the WD's atmosphere from accreted planetary debris. The largest category is by far the last, which includes over 1,000 systems \citep[][]{Dufour2007,Kleinman2013,Kepler2015,Kepler2016,Kepler2021,Coutu2019}. The smallest category is the first, with just five examples of major planets known \citep[][]{Thorsett1993,Sigurdsson2003,Luhman2011,Gaensicke2019,Vanderburg2020,Blackman2021}. Nevertheless, both categories are connected, despite this observational gulf, because planets can dynamically drive this debris or their progenitor asteroids, moons or comets into WDs.

The accreted debris is ubiquitous, appearing in between 25 and 50 per cent of Milky Way WDs \citep[][]{Zuckerman2003,Zuckerman2010,Koester2014}. The debris has also been observed to occur at cooling ages (the time since becoming a WD) up to 8 Gyr \citep[][]{Hollands2018,Blouin2022}.

Identifying the planetary architectures which can allow planets to perturb smaller bodies towards the WD in a manner that mimics the distribution of accretion rate with cooling age is an ongoing challenge and is subject to a large number of degeneracies. The currently known exoplanets orbiting main-sequence stars will predominately be engulfed by their parent stars upon leaving the main sequence \citep{Maldonado2020a,Maldonado2020b,Maldonado2021}, and sub-Saturn sized planets remain largely hidden from view at separations where such planets would survive.

However, such distant super-Earths is a class of planets shown to be particularly efficient at polluting WDs at a wide variety of cooling ages \citep{Frewen2014,Mustill2018}, particularly because such planets meander with their orbits in continuous motion \citep{Veras2015,Veras2016}. In contrast, a Jupiter-mass planet acts more like a sledgehammer on a fixed orbit, dissipating the system and polluting the WD in short bursts \citep{Veras2021b}. One key feature of these super-Earth polluters is that they are on eccentric orbits \citep[as e.g. for $\beta$ Pictoris;][]{Beust1996}, because a single planet on an exactly circular cannot perturb minor bodies sufficiently close towards a WD \citep{Antoniadou2016}. Also helpful for pollution is when planets reside in multiple-planet systems, because otherwise perturbing minor bodies onto star-grazing orbits is challenging \citep{Bonsor2011} and may require asteroid reservoirs several orders of magnitude more massive than the Solar Systems' \citep{Debes2012}. WD pollution due to secular chaos in multi-planet systems has been previously investigated in \cite{Smallwood2018,Smallwood2021} and \cite{Connor2021}.

The vast majority of WD pollution investigations which contain perturbing planets (see Fig. 6 of \citealt*{Veras2021} for an extensive list) use initial conditions for their planetary systems which are not outputs from birth cluster simulations. In a first attempt to bridge this gap, \cite{Veras2020} connected the outcomes of stellar cluster simulations involving outer solar system analogues with their future evolution across different stellar phases. However, partly because their setup was limited to giant planets on nearly circular orbits, that architecture is not necessarily representative of those found in chemically polluted WD systems\footnote{The fate of the Sun itself appears to be one of a polluted white dwarf \citep{Li2022}, which helps to highlight the importance of considering different reservoirs of minor body material when evolving these Solar system analogues \citep{Veras2020}.}.

Here, we perform stellar cluster simulations with a wider variety of planetary architectures which are more likely to pollute the eventual WDs over long cooling times. We also use more representative progenitor WD masses ($1.5\,\mathrm{M}_{\odot} - 2.5\,\mathrm{M}_{\odot}$; \citealt*{Tremblay2016,Cummings2018,McCleery2020}) rather than $1.0\,\mathrm{M}_{\odot}$ Sun-like stars. A key result of this study is publicly-available sets of post-cluster initial conditions that modellers could use as starting points for their simulations of post-main-sequence planetary systems.

Given the computational expense and complexity of stellar cluster simulations which contain multi-planet systems, we devote Section 2 towards describing our methods. In Section 3 we report the results, and we state our conclusions in Section 4. Our output data tables are available as supplementary material in the online version of this paper.

\section{Methods and Initial Conditions}

\subsection{Computational Approach}
Stars form predominantly in groups, like stellar associations or star clusters \citep[][]{Lada2003, PortegiesZwart2010}, and due to the close connection between star and planet formation, planets are accordingly born into these clustered environments. However, the simulation of multi-planetary systems in star clusters is challenging due to various reasons and requires different computational approaches depending on the underlying scientific question. One challenge for the numerical integration of the motions of the planets around the stars and the motion of the host star through the cluster are the completely different dynamical timescales. While the dynamical evolution of planets takes place on timescales of days and years, for star clusters it is typically in the range of several million years. Another aspect is the hierarchical nature of stars with (multi-)planetary systems. In principle, planetary systems can be treated and regularized similar to binary systems. \cite{Spurzem2009}, who studied single-planetary systems in a star cluster in a fully coupled dynamical simulation, used this approach, as well as \cite{vanElteren2019}, who studied multi-planetary systems in star clusters using the \texttt{Nemesis} module in \texttt{AMUSE} \citep[][]{PortegiesZwart2011, AMUSEbook}.
However, we want to be able to accurately trace resonant and secular effects in the dynamical evolution of the planetary systems. For this reason, we use a hybrid approach and simulate star cluster and planetary systems separately by using encounter information from the star cluster simulation for the integration of the planets. This approach is possible under the assumption that the motion of individual stars and the evolution of the whole star cluster can influence the dynamical evolution of the planets, but not vice versa.

As a first step in this approach, the star cluster is simulated using \texttt{NBODY6++GPU} \citep[][and references therein]{Aarseth2003,Wang2015b,Wang2016}, where the motions of the stars are integrated using the Hermite scheme \citep[e.g.][]{Aarseth2003, Aarseth2008}. All necessary information is stored in high temporal resolution in the \texttt{HDF5}\footnote{\url{https://www.hdfgroup.org/}} format. Then, using the \texttt{LonelyPlanets Scheme} \citep[\texttt{LPS};][]{Cai2015, Cai2017, Cai2018, Cai2019, FlamminiDotti2019, Stock2020}, which is based on the \texttt{AMUSE} framework, all encounters of the selected host stars with each of the five nearest stars during the cluster simulation are calculated and stored, including the first and second time derivatives of the perturbers, in order to reconstruct the details of an encounter for the subsequent integration of the planetary systems. \texttt{LPS} uses \texttt{REBOUND} \citep[][]{Rein2012} for the actual integration of the planetary systems, as well as additional features from \texttt{REBOUNDx} \citep[][]{Tamayo2020a}. For our simulations, we use \texttt{REBOUND}'s high-order, adaptive-step size integrator \texttt{IAS15} \citep[][]{Rein2015} to obtain accurate  integration results of systems with close encounters between the planets.

\subsection{Initial Conditions for the Star Cluster Simulation}\label{sec:IC_cluster}

We simulate an open star cluster consisting of 8000 stars whose masses follow a \cite{Kroupa2001} initial mass function (IMF) in the mass range of $0.08-20\,M_\odot$. The total cluster mass is $M_\mathrm{cl} = 4073.4\,\mathrm{M}_\odot$. The star's initial positions and velocities in the cluster are drawn from a \cite{Plummer1911} model. The star cluster is initially in virial equilibrium ($|U| = 2T$, where $U$ is the total potential energy of the Plummer sphere and $T$ is the total kinetic energy of the cluster stars). The virial radius, which is defined as $r_\mathrm{vir}= \mathrm{G}M_\mathrm{cl}^2/(2|U|)$ (with $\mathrm{G}$ being the gravitational constant), is $1\,\mathrm{pc}$ for our cluster, while the initial half-mass radius is $r_\mathrm{hm} \approx 0.78\,\mathrm{pc}$. 
The star cluster is assumed to be on a Solar-like orbit around the Galactic centre which is why the tidal forces of the Galaxy acting on the cluster are assumed to be equal as for the solar neighbourhood \citep{Heisler1986}.
The cluster's initial tidal radius $r_\mathrm{tid}=R_\mathrm{G}(M_\mathrm{cl} / M_\mathrm{G})^{1/3}$ (with $R_\mathrm{G}$ being the distance to the Galactic centre and $M_\mathrm{G}$ being the Galaxy's mass contained inside $R_\mathrm{G}$) is $22.6\,\mathrm{pc}$. Stellar evolution is implemented \citep[see e.g.][]{Spera2015, Khalaj2015}, but the mass loss is negligible for the host stars whose planetary systems are only simulated during a period when their host star is still on the main sequence.
Primordial binary systems are not included, however, binaries can form during the course of the simulation. We also do not assume primordial mass segregation, but we observe the onset of mass segregation during the first few million years when the cluster experiences a short phase of core collapse. 

According to equation 3 in \cite{Malmberg2007}, encounters below $r_\mathrm{min}=1000\,\mathrm{au}$ between our host stars and an average-massed star ($M_\star \sim 0.51\,\mathrm{M}_\odot$) in the cluster take place on timescales of $\tau_\mathrm{enc}\approx$ 0.6 -- 1.2$\,\mathrm{Myr}$. This corresponds to encounter rates of 0.8 -- 1.7 encounters per star and per Myr for the host-star mass range used in our simulation.

\begin{figure*}
    \centering
    \includegraphics[width = 0.49\textwidth]{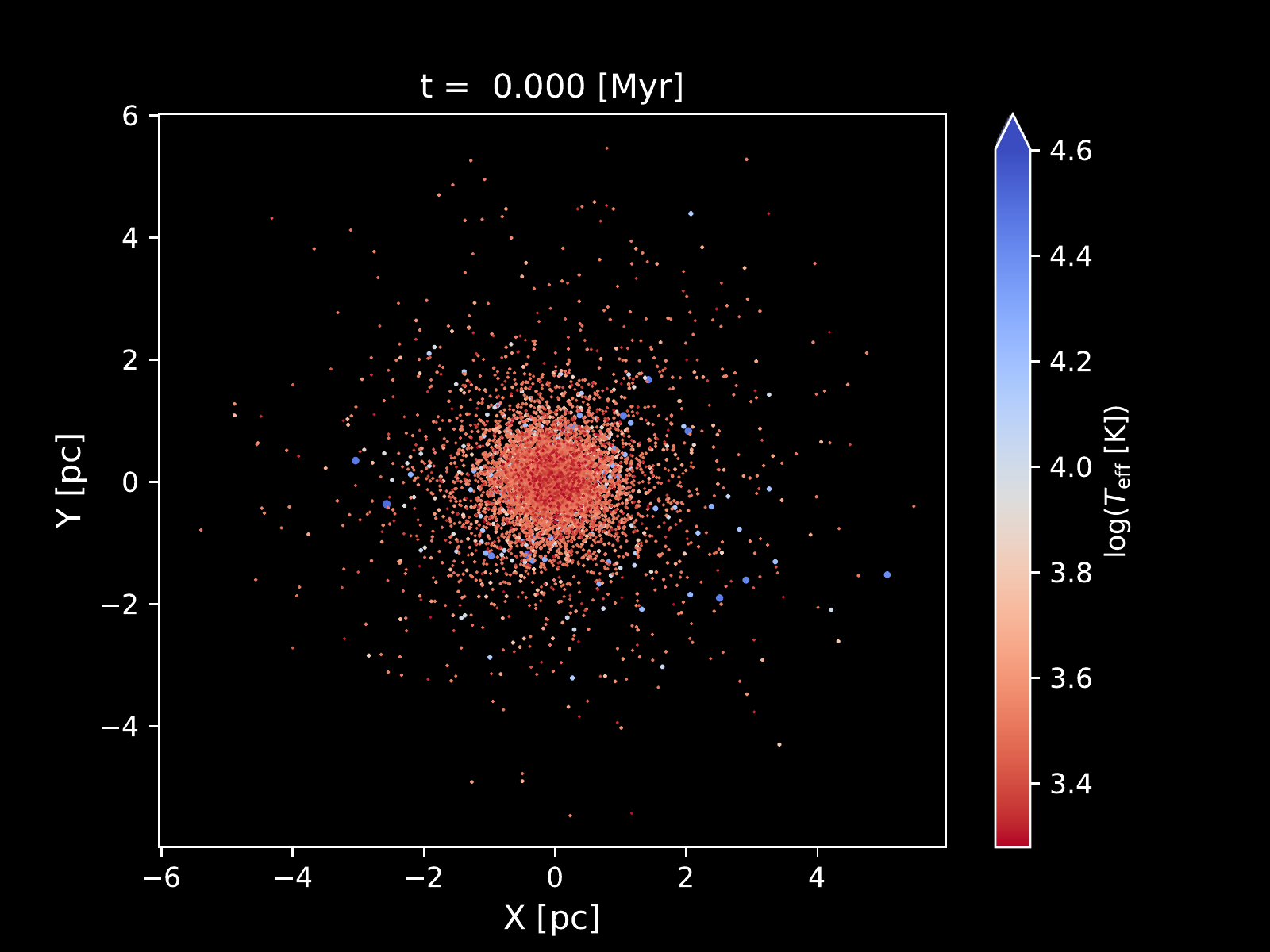}
    \includegraphics[width = 0.49\textwidth]{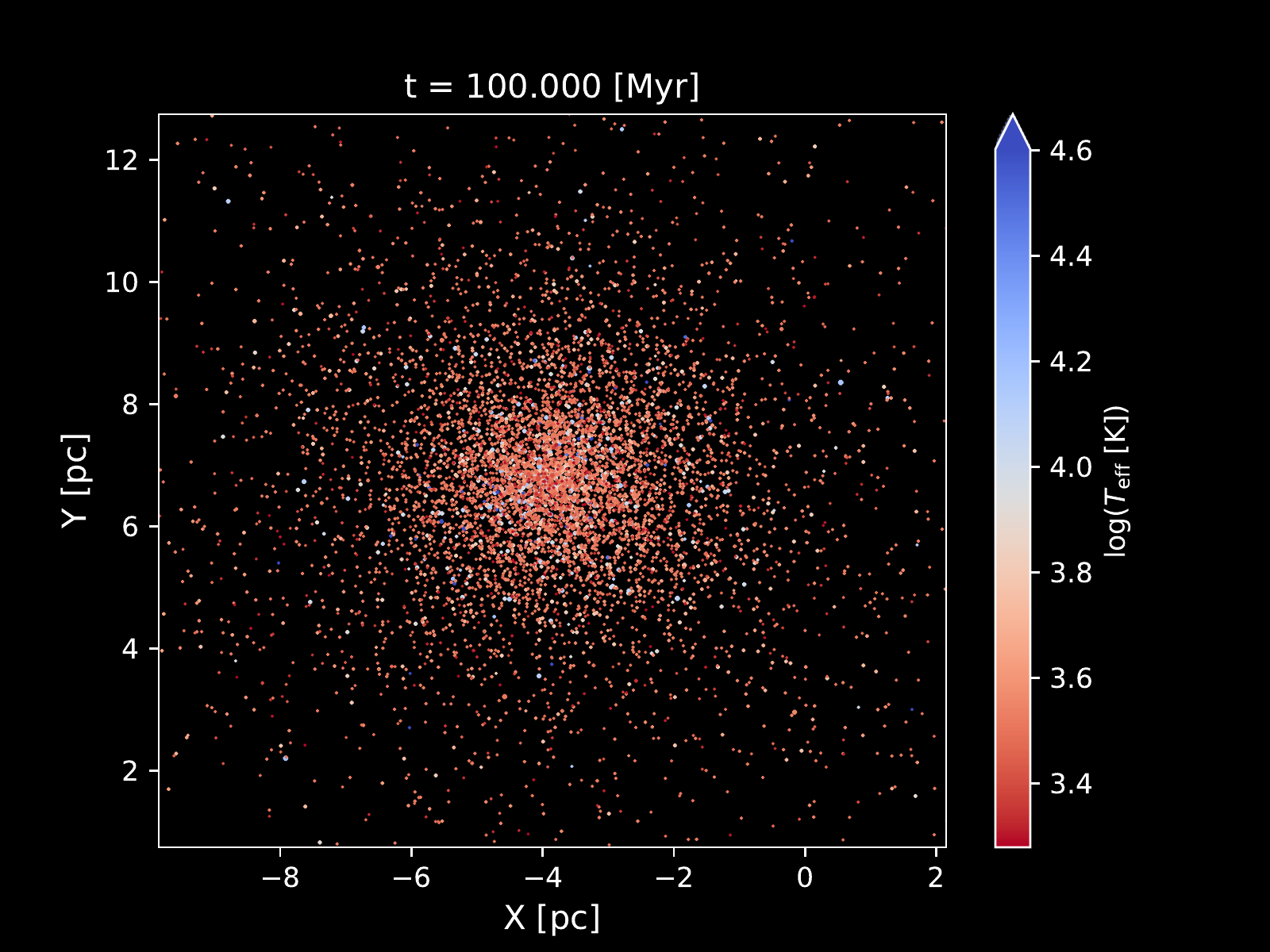}
    \caption{The simulated star cluster at the beginning of the simulation ($t=0\,\mathrm{Myr}$, $r_\mathrm{vir}=1.0\,\mathrm{pc}$; left panel) and at the time $t=100\,\mathrm{Myr}$ ($r_\mathrm{vir}=3.9\,\mathrm{pc}$; right panel). The star cluster has visibly expanded and is in the process of dissolving. Although the cluster's centre of mass and its density centre moves, which means that the plotted axes range must be adjusted with increasing simulation time, the physical scale (a total of $12\,\mathrm{pc}$ on x- and y-axis) for both plots remains the same.}
    \label{fig:star_cluster_evolution}
\end{figure*}

We want to integrate the planetary systems until the cluster has sufficiently dissolved. Here, however, a compromise must be found between the computational costs of the planetary system simulations and the complete dissolution of the cluster. A good compromise for the 8000 star cluster we simulate is a period of $100\,\mathrm{Myr}$. Although only about 20 per cent of the stars have completely escaped the cluster's gravitational field by then, the cluster has already expanded significantly to $r_\mathrm{vir} = 3.87\,\mathrm{pc}$, so that encounters between stars outside the dense core are very rare after more than $100\,\mathrm{Myr}$.
The median distance of the host stars to the cluster centre after $100\,\mathrm{Myr}$ is around $8\,\mathrm{pc}$ for the lowest and highest host star mass range, while the median distance to the nearest star is $1.1\,\mathrm{pc}$ and $1.4\,\mathrm{pc}$, respectively, and has increased by a factor of 11 and a factor of 15 compared to the beginning of the simulation. Thus a maximum integration time of $100\,\mathrm{Myr}$ for the planetary systems seems to be adequate. An optical comparison between the star cluster at the beginning of the simulation and after $100\,\mathrm{Myr}$ is shown in Fig.~\ref{fig:star_cluster_evolution}.

\subsection{Initial Conditions for the Planetary System Simulations}\label{sec:ic_planets}
In this work we provide the results of 1224 planetary systems embedded in an open star cluster whose properties were described in Sec.~\ref{sec:IC_cluster}. The host stars to be simulated are the typical progenitor stars of polluted WDs on the main sequence, which typically have masses of $1.5$ -- $2.5\,\mathrm{M}_\odot$ \citep[see, for example,][]{Tremblay2016,Cummings2018,ElBadry2018,McCleery2020,Barrientos2021}. 

However, using a continuous mass spectrum for the host stars, as would be the case in a real star cluster, would reduce the comparability between the individual simulated planetary systems. For this reason, we divide the host stars into three different mass ranges, $1.25$ -- $1.75\,\mathrm{M}_\odot$, $1.75$ -- $2.25\,\mathrm{M}_\odot$ and $2.25$ -- $3.25\,\mathrm{M}_\odot$, and search for those stars whose masses lie in one of these mass ranges. Since the IMF drops off very steeply towards higher masses and the number of available stars is limited in the range around $2.5\,\mathrm{M}_\odot$, the upper limit of the third mass range is deliberately chosen to be higher. 
We then calculate the encounters of all these stars with each of the nearest five stars in the cluster and store this information. For the subsequent integration of the planetary systems, the masses of the host stars are set to $1.5\,\mathrm{M}_\odot$, $2.0\,\mathrm{M}_\odot$ and $2.5\,\mathrm{M}_\odot$ to ensure the comparability of the planetary systems within these three mass ranges and to be able to work out the pure effect of the cluster environment on the dynamical evolution of the individual systems. According to the number of stars present in the three mass ranges, we have a total of 408 host stars available (193 stars with $M_\star = 1.5\,\mathrm{M}_\odot$, 114 stars with $M_\star = 2.0\,\mathrm{M}_\odot$ and 101 stars with $M_\star = 2.5\,\mathrm{M}_\odot$).

The planetary systems around these 408 host stars are then started in three different initial orbital configurations, while the host star and its trajectory through the cluster remain the same for all three different planetary system models. All planetary systems solely consist of super-Earths, each having a mass of 0.01~$\mathrm{M}_\mathrm{Jup}$ ($\approx$ 3.2~$\mathrm{M}_\oplus$). Due to the variety in multiplicity of actual observed planetary systems, we aim to simulate two bounding cases of systems consisting of three and seven planets. However, the compactness of a 7-planet system crucially determines its dynamical evolution, especially if it is externally perturbed (by stellar flybys).

Therefore, we simulate the following three scenarios in which all planets are equally separated in terms of mutual Hill radii ($R_\mathrm{H,m} = (a_1+a_2)/2 \cdot \big((M_1+M_2)/(3M_\star)\big)^{1/3}$): (i) a system consisting of only three planets (called ``3P model'') between 2.0 and 18.63~au, (ii) a rather tightly packed system of seven planets, called ``7PC model'', within the same orbital boundaries as model 3P, and (iii) a wider system of seven planets, called ``7PW model'', in which the five innermost planets are placed in the same orbital range as in the previous two cases, with two additional planets on wider orbits (the outermost planet has $a=56.87\,\mathrm{au}$, resulting from the fixed number of mutual Hill radii).

The orbital configurations of the three models and the number of mutual Hill radii used for the orbital spacing are listed in Table~\ref{tab:IC_planetary_systems} and Table~\ref{tab:Hill_radii}, respectively. In all three models the planetary orbits are initially circular ($e=0$) and coplanar ($i=0^\circ$).  Furthermore, all systems are long-term stable if they are placed in isolation. The argument that the planetary systems should be stable in isolation over time was also decisive for outwardly increasing spacings between the planets, which we achieve by using mutual Hill radii instead of using similar orbital spacings according to the peas-in-a-pod theory \citep[e.g.][]{Millholland2017, Weiss2018} based on findings from the Kepler mission.

The inner boundary of 2.0~au is chosen as a minimum semimajor axis to account for the potential engulfment of the innermost planet due to the expansion of the host star during the giant branch phase. As a basis for the outer boundary in the first two cases, we take into account that core accretion during planet formation becomes inefficient at larger semimajor axes, although it is not impossible for super-Earths to form at wider orbits \citep[see, for example, fig.~D.3 in][]{Schlecker2021} which is why we additionally consider the possibility of a more extended planetary system in the third scenario.

\begin{table}
    \centering
    \caption{The planet's initial semimajor axes (in au) for the three different planetary system models. All orbits are initially circular, coplanar and aligned. The mass of each individual planet is $M_\mathrm{pl}=0.01\,\mathrm{M}_\mathrm{Jup}$.}
    \begin{tabular}{l|ccccccc}
        \hline
         Model & P1 & P2 & P3 & P4 & P5 & P6 & P7\\
         \hline
         3P & 2.00 & 6.10 & 18.63 \\
         7PC & 2.00 & 2.90 & 4.21 & 6.10 & 8.86 & 12.85 & 18.63 \\
         7PW & 2.00 & 3.49 & 6.10 & 10.67 & 18.63 & 32.55 & 56.87 \\
         \hline
    \end{tabular}
    \label{tab:IC_planetary_systems}
\end{table}

\begin{table}
    \centering
    \caption{\ks{The number of mutual Hill radii $R_\mathrm{H,m}$ between the planets in each model.}}
    \begin{tabular}{l|ccc}
        \hline
         \ks{Model} & \ks{$1.5\,\mathrm{M_\odot}$} & \ks{$2.0\,\mathrm{M_\odot}$} & \ks{$2.5\,\mathrm{M_\odot}$}\\
         \hline
         \ks{3P} & \ks{62.6} & \ks{68.9} & \ks{74.2}\\
         \ks{7PC} & \ks{22.7} & \ks{25.0} & \ks{26.9}\\
         \ks{7PW} & \ks{33.6} & \ks{37.0} & \ks{39.8}\\
         \hline
    \end{tabular}
    \label{tab:Hill_radii}
\end{table}

\section{Results and Discussion}

\subsection{Fraction of Surviving Planets}
If the eccentricity of a planet is excited to more than $e > 0.99$, we assume that the planet is close to ejection and remove it from the simulation. These ejected planets would, depending on their escape velocity, either continue to move through the cluster as free-floating planets, which may even allow re-capture by other cluster members, or they would directly escape not only the gravitational field of the host star, but also that of the star cluster. 

\begin{figure}
    \centering
    \includegraphics[width = 0.49\textwidth,trim= 0.7cm 0.2cm 1.1cm 0.0cm,clip]{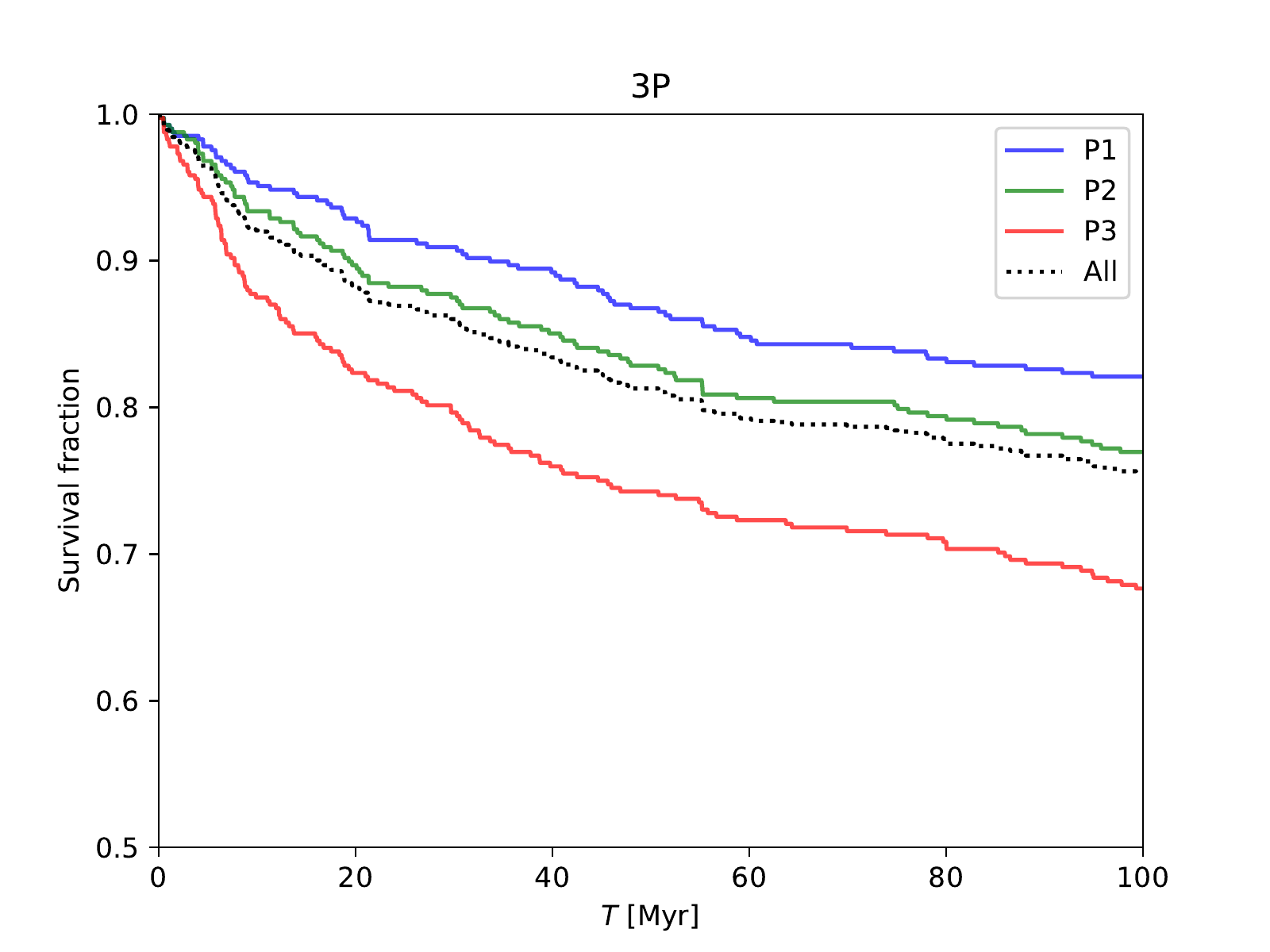}
    \includegraphics[width = 0.49\textwidth,trim= 0.7cm 0.2cm 1.1cm 0.0cm,clip]{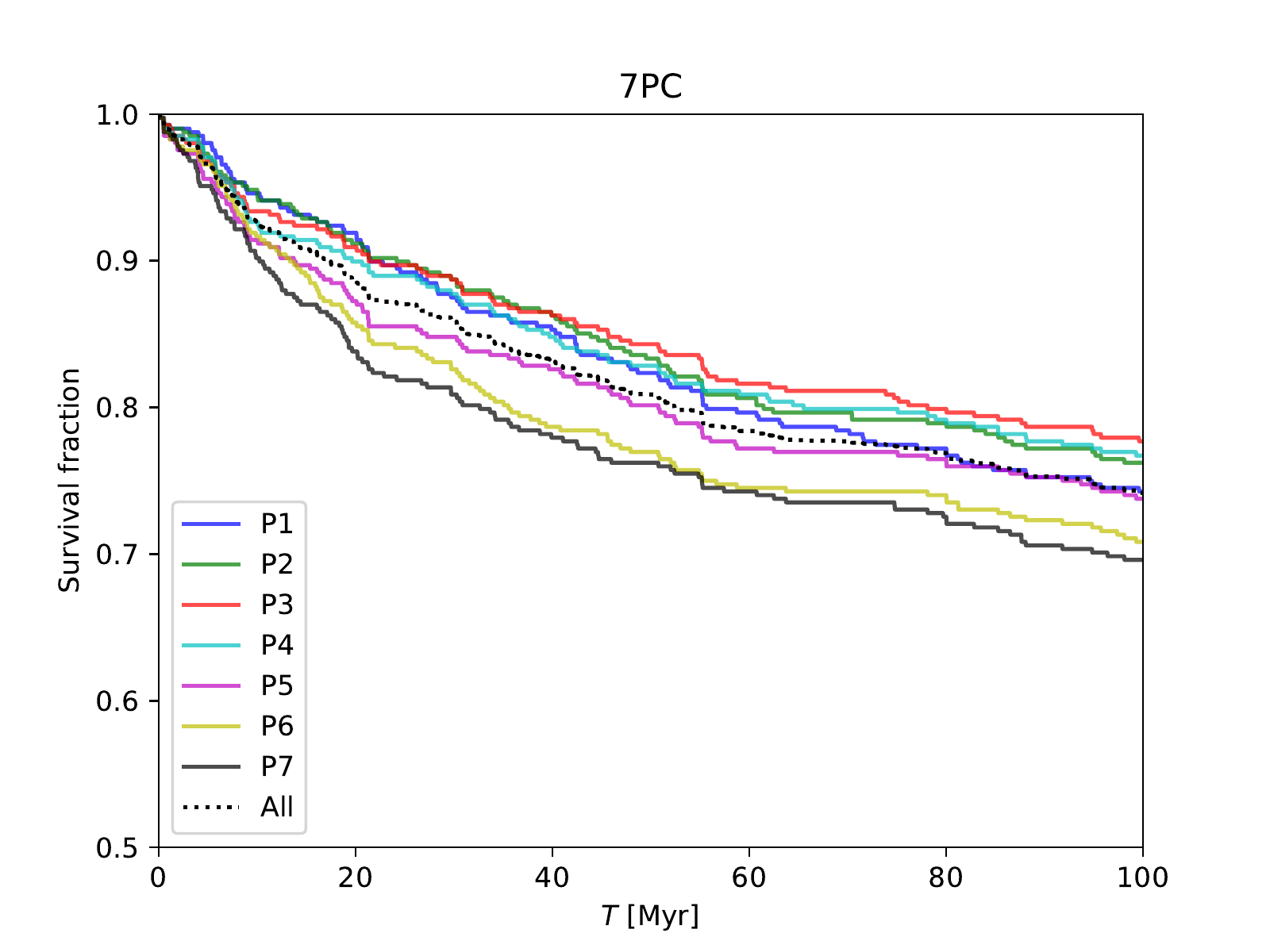}
    \includegraphics[width = 0.49\textwidth,trim= 0.7cm 0.2cm 1.1cm 0.0cm,clip]{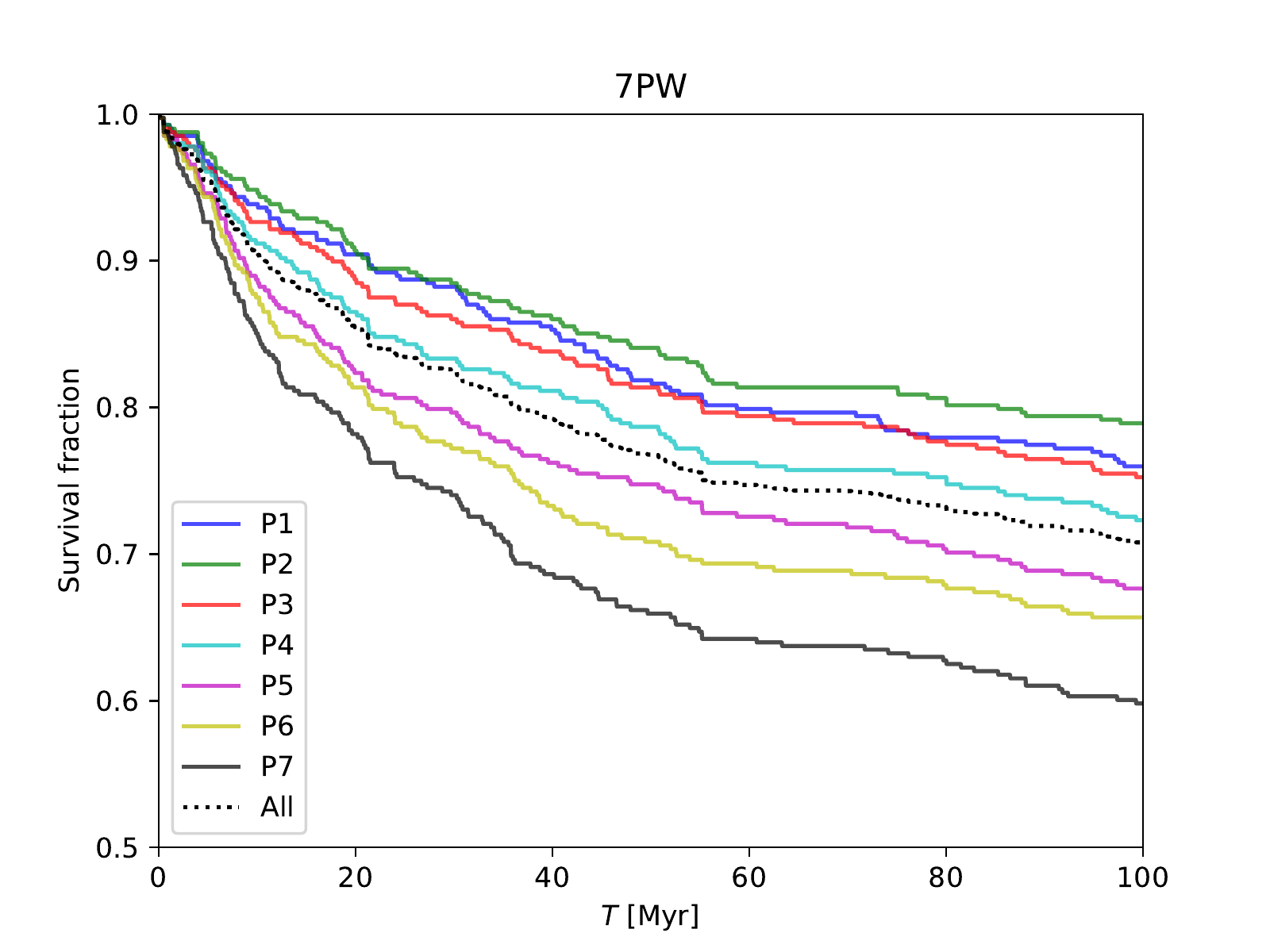}
    \caption{Fractions of surviving planets for the 3P, 7PC, and 7PW model. The dotted black line represents the average survival fraction for each model.}
    \label{fig:survival_rates}
\end{figure}

For technical reasons, we do not trace the motion of the planets through the cluster after their ejection from a planetary system. However, the fraction of ejected planets $f_\mathrm{ej}$ gives an estimate for the expected number of free-floating planets in open clusters with similar properties to the one we simulated. The fraction of surviving planets $f_\mathrm{surv}= 1- f_\mathrm{ej}$ is plotted in Fig.~\ref{fig:survival_rates} for all three planetary models as a function of simulation time. The model with the highest $\overline{f}_\mathrm{surv}$ (dotted black line in Fig.~\ref{fig:survival_rates}), averaged over all planets in the system, is the 3P model with a value of $0.76$. The 7PC model shows little difference with $\overline{f}_\mathrm{surv} = 0.74$, indicating that despite the higher planet density in this system compared to the 3P model, which in principle leads to more planet-planet interaction and to higher ejection rates, the orbit width of the outer planets is the more important factor for the averaged survival fraction. Consequently, the planets in the 7PW model have on average the lowest survival probability with a value of $\overline{f}_\mathrm{surv.} = 0.71$. The fraction of escapers that arise in our simulations are slightly higher than e.g. in \cite{vanElteren2019}, who obtained $\overline{f}_\mathrm{ej} \approx 0.14$. However, the difference can be explained by the significantly shorter simulation time and the considerably smaller number of stars in the host star cluster in \cite{vanElteren2019}. 

However, when considering the survival probability for the individual planets, the exact configuration of the planetary system, especially its multiplicity and consequently its compactness, does play a role. While in the 3P model the planet's probability for being ejected correlates with its initial semimajor axis, this is not consistent with the 7PC and 7PW models, as can be seen in Fig.~\ref{fig:survival_rates}.

The planet with the highest survival fraction in the 7PC model is P3 ($f_\mathrm{surv}= 0.78$), followed by P4 ($f_\mathrm{surv}=0.77$), P2 ($f_\mathrm{surv}=0.76$) and P1 ($f_\mathrm{surv}=0.74$), so the survival rate increases slightly for the middle planets and only decreases from the fifth planet towards the outer planets P7 (which has $f_\mathrm{surv}=0.70$).
The spread in survivability for the 3P model is only slightly larger and ranges from $f_\mathrm{surv} = 0.68$ to $f_\mathrm{surv} = 0.82$. The reason for the changed order in the survival fraction of the planets in the 7PC model is the higher planetary density at constant orbital expansion of the system. Due to increased interactions among the planets after an external gravitational perturbation, the inner planets can experience delayed ejection from the system indirectly as a result of an earlier flyby of a neighbouring star. These delayed ejections have also been observed in \cite{vanElteren2019} and \cite{Stock2020}.

As expected, for a system with wider orbits but the same number of planets, as in the 7PW model, the spread in the individual planet's survival rate is larger than for the more compact case. Here the values are between $f_\mathrm{surv}=0.60$ (P7) and $f_\mathrm{surv}=0.79$ (P2). As in the 7PC model, the second innermost planet in the 7PW model has a slightly larger survival fraction than the innermost planet P1 ($f_\mathrm{surv}=0.76$), but here the planets' survival probability decreases beyond the second planet as expected with increasing initial semimajor axes.

\subsection{Semimajor Axis and Eccentricity Distribution and Possible Engulfment during Red Giant Phase}

\begin{figure*}
    \centering
    \includegraphics[width = 0.49\textwidth,trim= 0.6cm 0.2cm 1.1cm 0.2cm,clip]{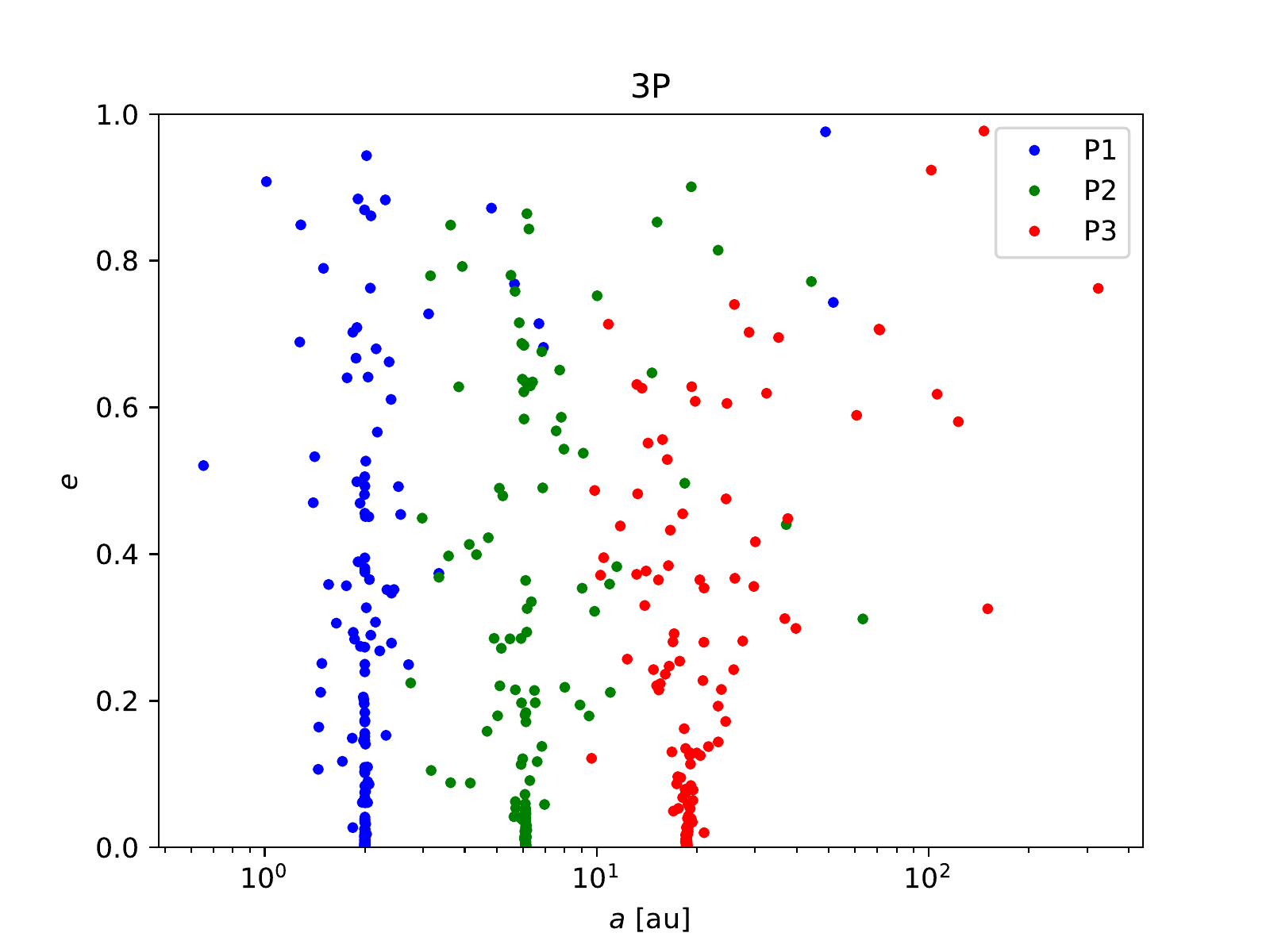}
    \includegraphics[width = 0.49\textwidth,trim= 0.6cm 0.2cm 1.1cm 0.2cm,clip]{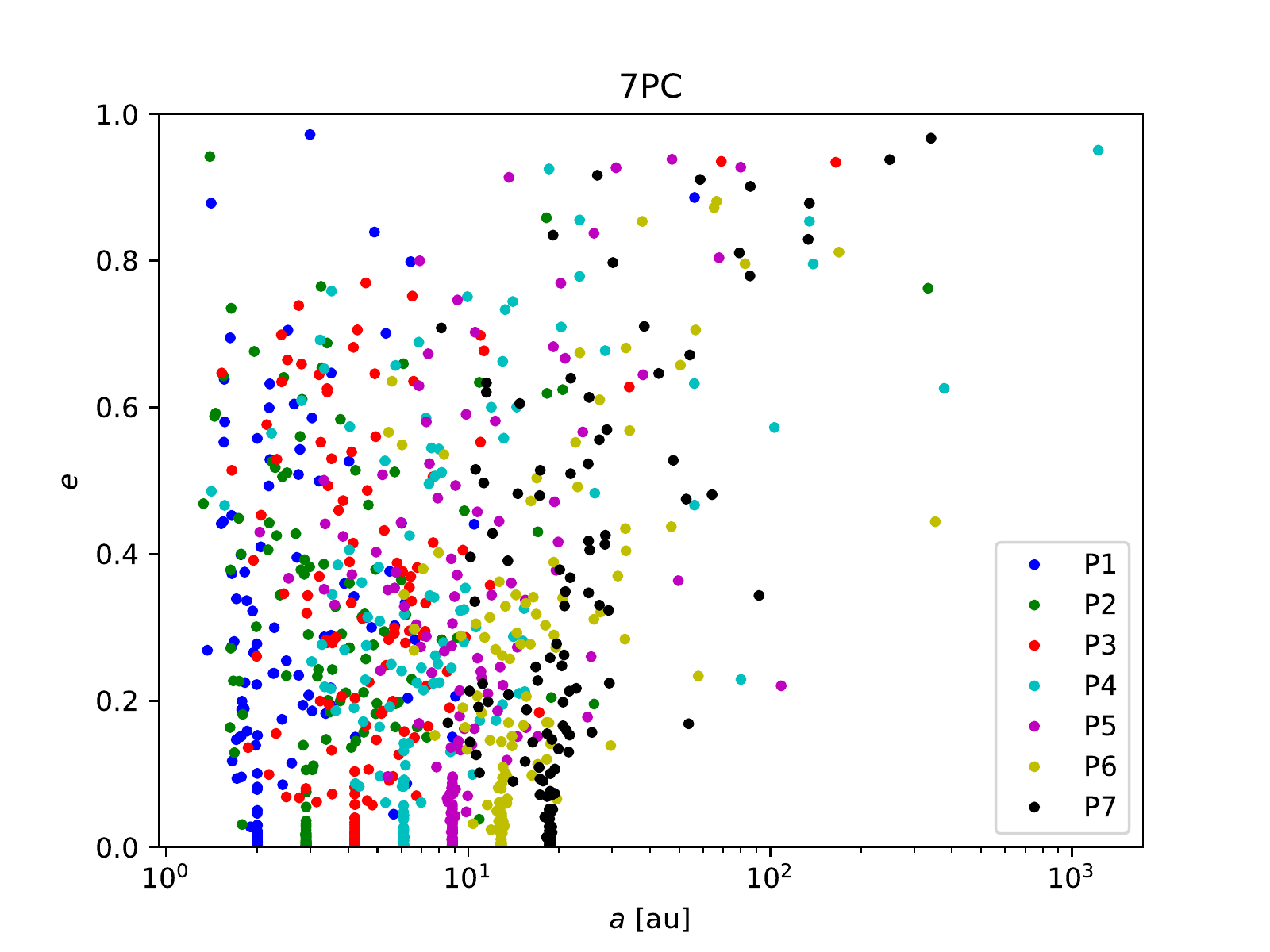}
    \includegraphics[width = 0.49\textwidth,trim= 0.6cm 0.2cm 1.1cm 0.2cm,clip]{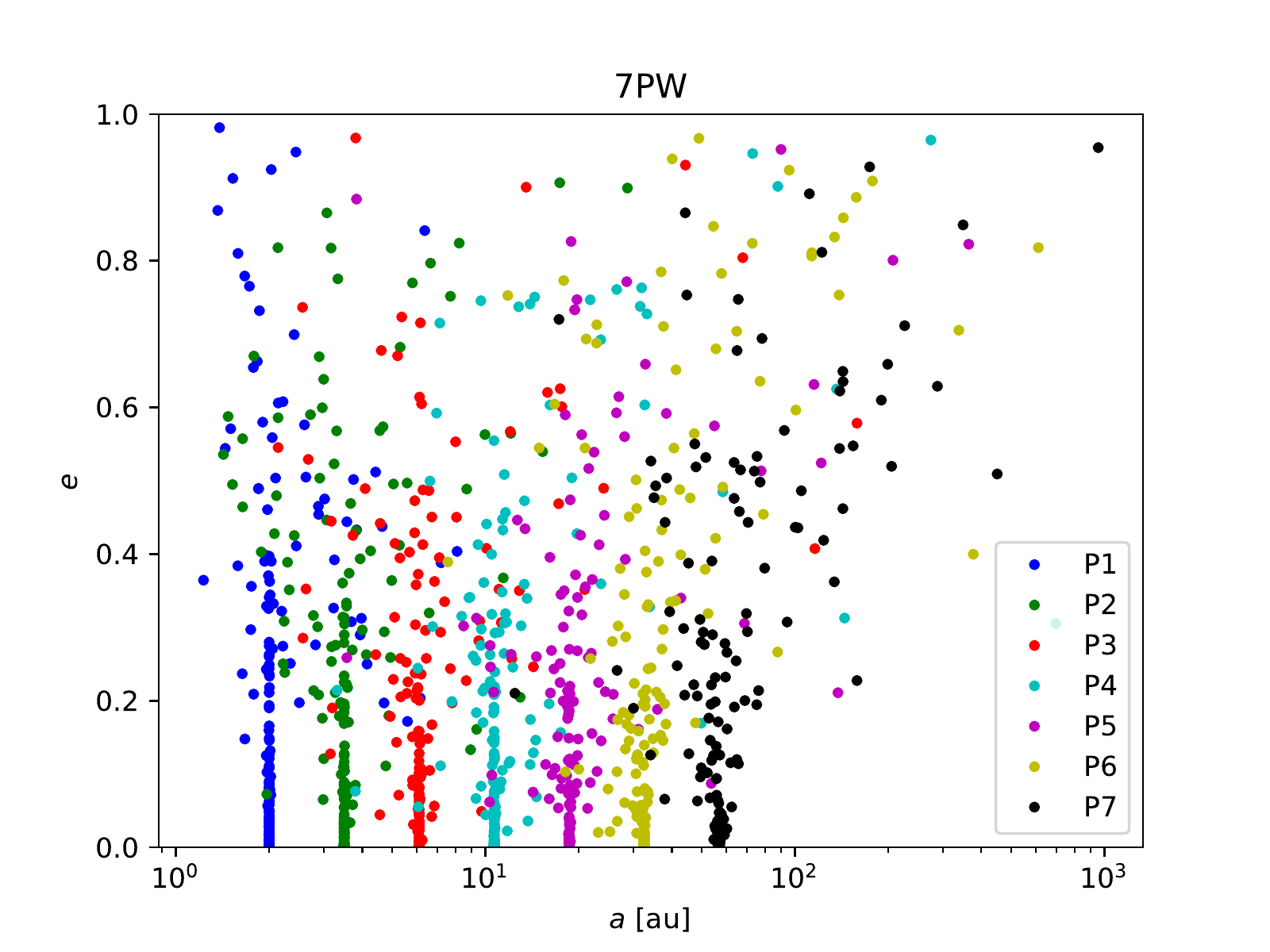}
    \includegraphics[width = 0.49\textwidth,trim= 0.6cm 0.2cm 1.1cm 0.2cm,clip]{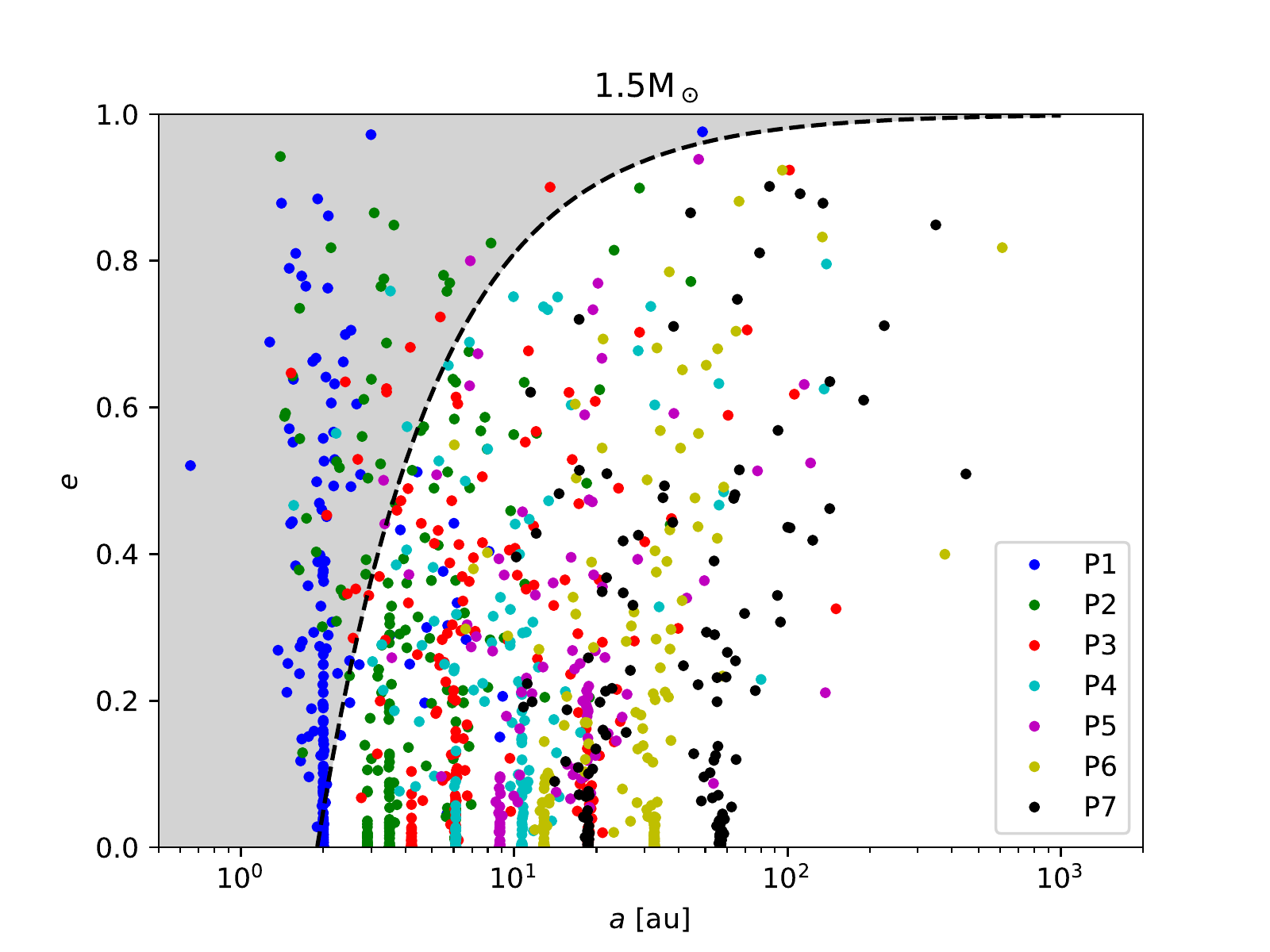}
    \includegraphics[width = 0.49\textwidth,trim= 0.6cm 0.2cm 1.1cm 0.2cm,clip]{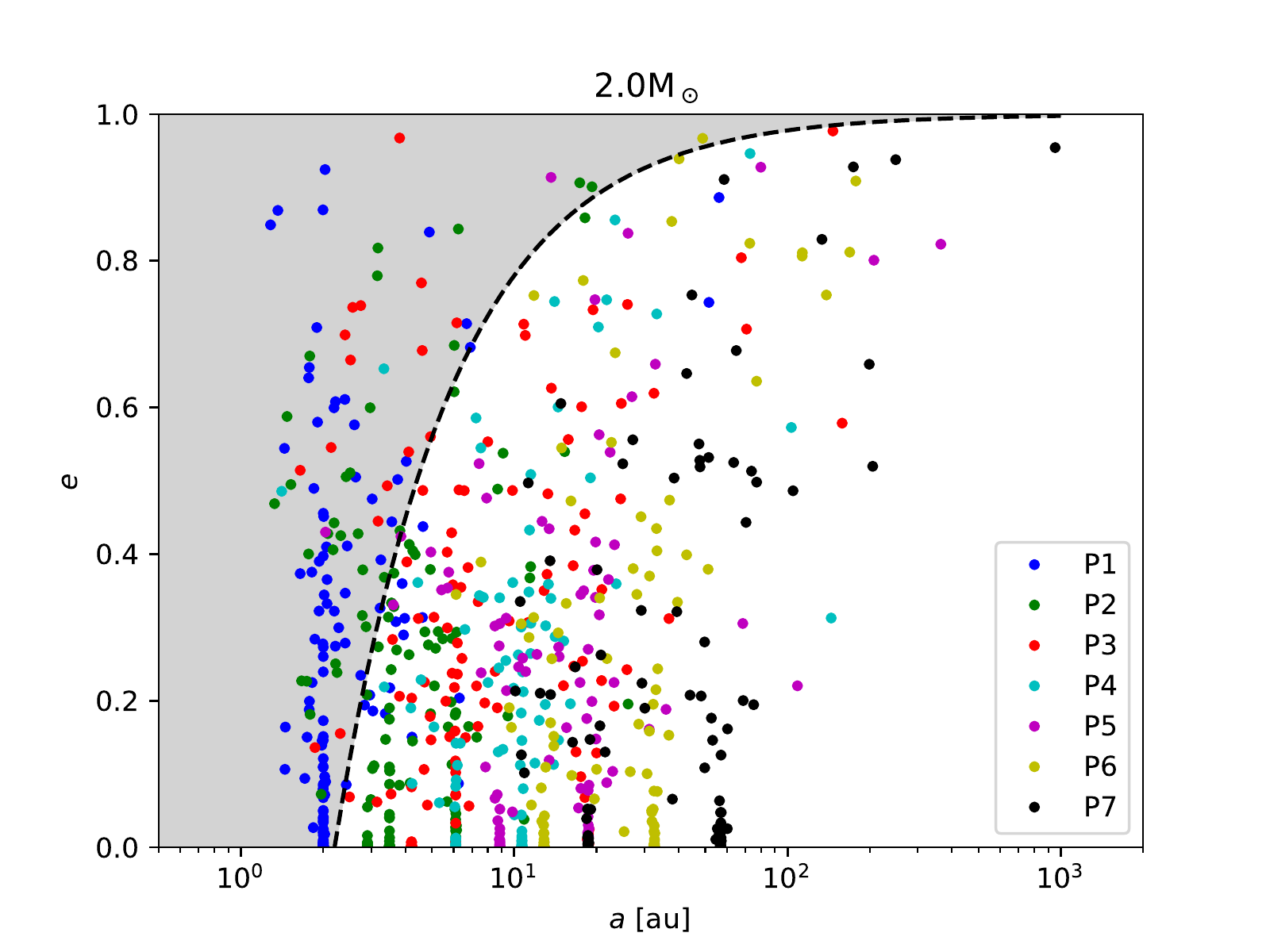}
    \includegraphics[width = 0.49\textwidth,trim= 0.6cm 0.2cm 1.1cm 0.2cm,clip]{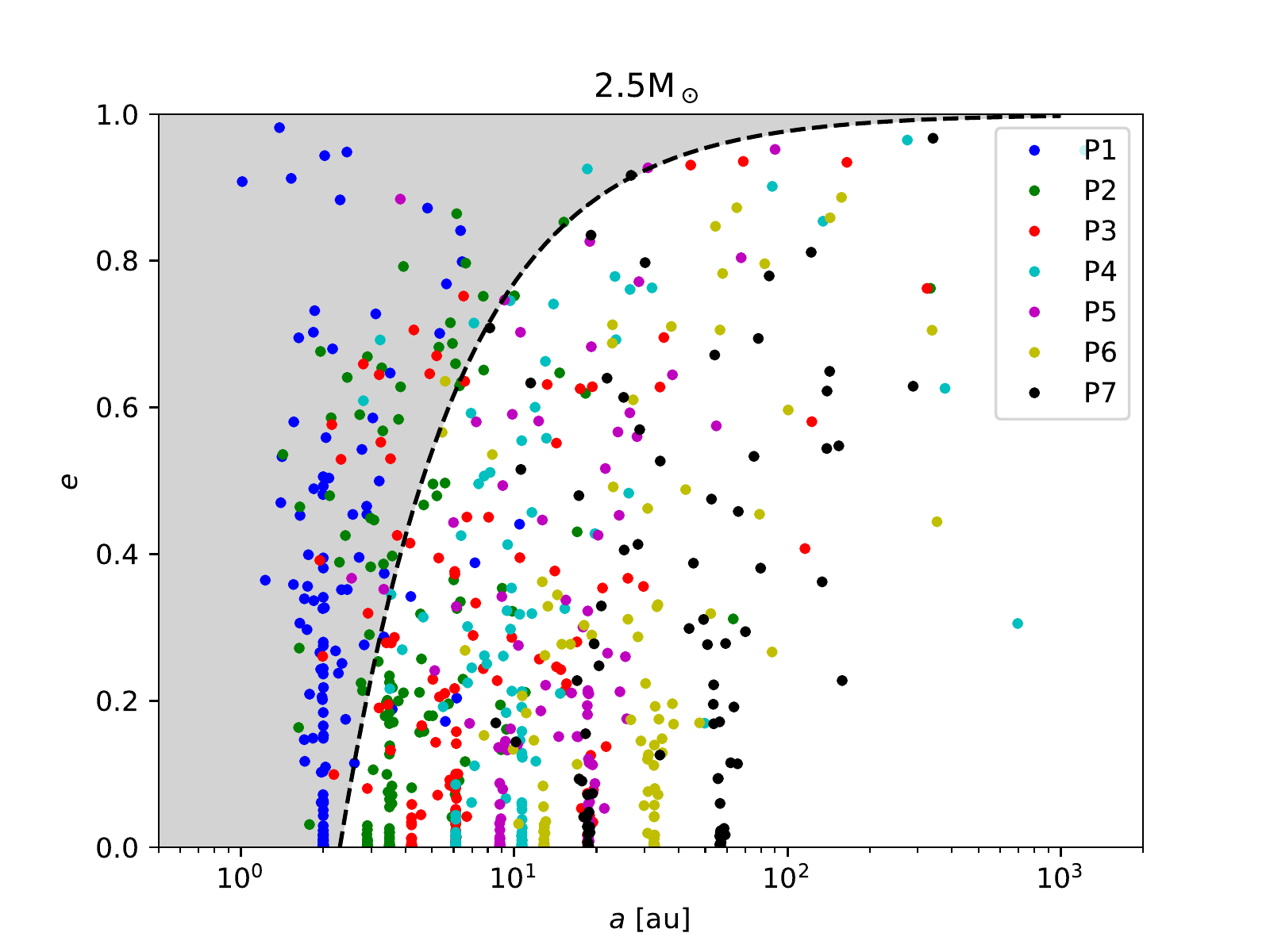}
    \caption{The $a$-$e$ space for the 3P, 7PC and 7PW model as well as for the different host star masses. The grey shaded area shows which planets would be engulfed by the star during the red giant phase. }
    \label{fig:a-e_space}
\end{figure*}

The fraction of planetary systems whose dynamical evolution is considerably perturbed by passing stars (directly or indirectly by delayed planet-planet scattering) depends on the one hand on the planetary model used, but also on the host star mass. For the 3P model and a host star mass of $1.5\,\mathrm{M}_\odot$, we generally observe the lowest effect of the stellar environment on the dynamical evolution of the individual planets. In this scenario, 83 per cent of the planets remain largely unperturbed. As a criterion for a considerable perturbation, we look at whether the semimajor axis deviates by more than 5 per cent from the initial value by the end of the simulation, or whether the eccentricity increases to more than 0.1. The fraction of planets that are significantly perturbed in their dynamics increases for the models with more planets per star, the orbital separation of the outermost planet, and the host star mass. For the 7PW model and a host star mass of $2.5\,\mathrm{M}_\odot$, the fraction of perturbed planets increases from 17 per cent to 29 per cent.
The distribution in $a$-$e$ space is wide for those planets that were considerably perturbed, as we demonstrate in Fig.~\ref{fig:a-e_space}, and comparable to results from previous studies \citep[see, e.g., fig. 10 in][]{Malmberg2011}. We observe inward migration for the perturbed planets in a few cases, but outward migration in the vast majority of cases. Almost 1 per cent of all planets have wide orbits of more than $100\,\mathrm{au}$ at the end of the simulation and one planet out of a total of 6936 planets is even scattered to an orbit of more than $1000\,\mathrm{au}$. More than 6 per cent of all planets are excited to high eccentricities ($e>0.5$). The fractions vary depending on the planetary system model and host star mass and are listed in Table~\ref{tab:fraction_orbits} for all the different scenarios.

\begin{table}
    \centering
    \begin{tabular}{c|cccc}
    \hline
    Model & $a > 100\,\mathrm{au}$ & $a > 1000\,\mathrm{au}$ & $e > 0.5$ & $i \ge 90^\circ$ \\
    \hline
    3P & 0.49 & 0.00 & 6.54 & 1.72\\
    7PC & 0.49 & 0.04 & 6.1 & 1.09\\
    7PW & 1.58 & 0.0 & 6.27 & 1.47\\
    $1.5\,\mathrm{M}_\odot$ & 0.67 & 0.00 & 4.94 & 1.31 \\
    $2.0\,\mathrm{M}_\odot$ & 0.98 & 0.00 & 6.24 & 1.29 \\
    $2.5\,\mathrm{M}_\odot$ & 1.40 & 0.06 & 8.68 & 1.51 \\
    \hline
    \end{tabular}
    \caption{Fraction of planets (in per cent) with wide ($a > 100\,\mathrm{au}$), very wide ($a > 1000\,\mathrm{au}$), very eccentric ($e > 0.5$) or retrograde ($i \ge 90^\circ$) orbits for the different planetary models (independent of the host star mass) and for the different host star masses (independent of the planetary system model used).}
    \label{tab:fraction_orbits}
\end{table}

\begin{figure}
    \centering
    \includegraphics[width= 0.47\textwidth,trim= 0.6cm 0.2cm 1.25cm 0.1cm,clip]{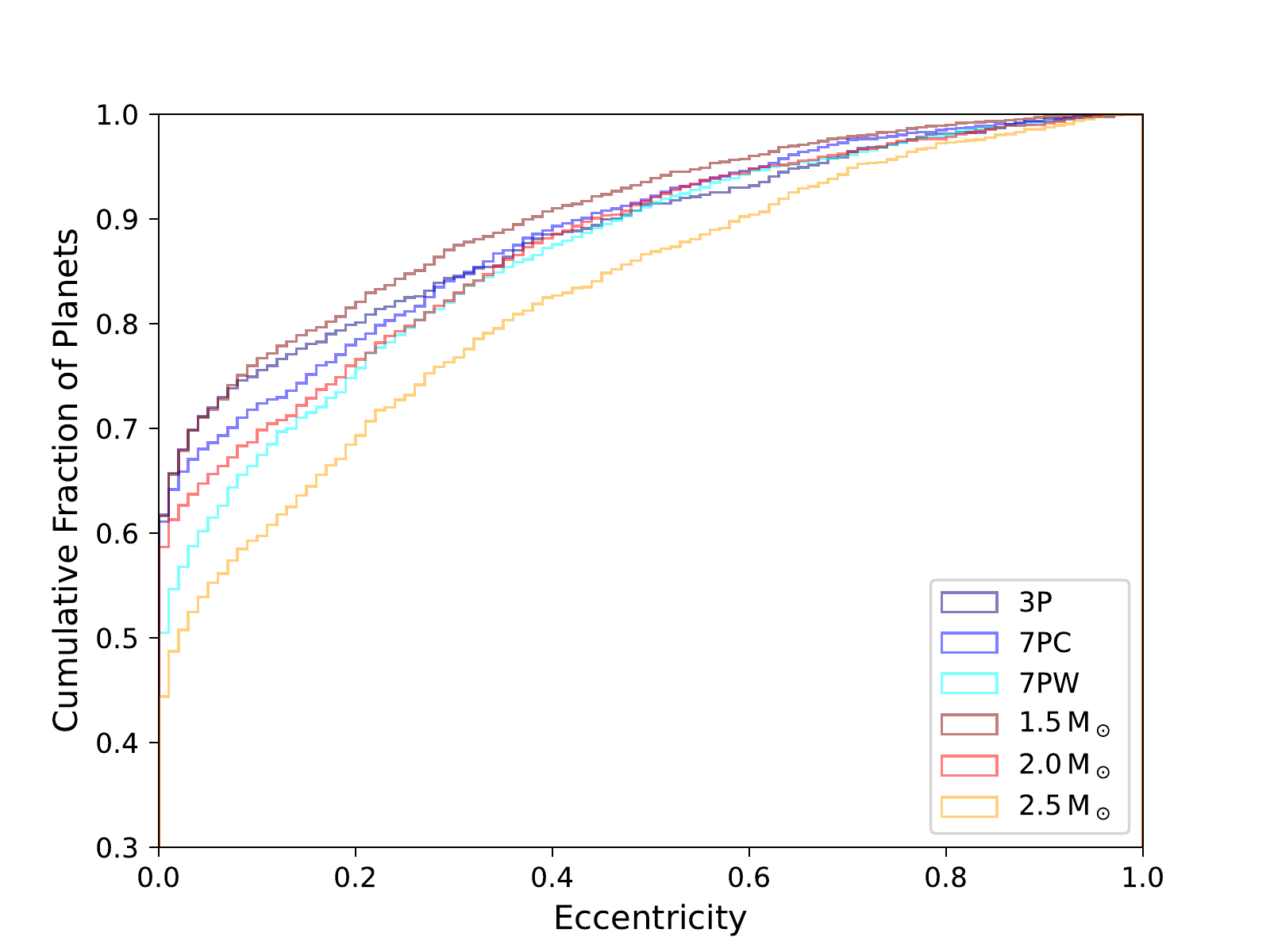}
    \caption{Cumulative, normalized histogram showing the distribution of eccentricities sorted by planetary model (bluish colors) and host star mass (reddish colors). The bin size is 0.02.}
    \label{fig:cumulative_hist_ecc}
\end{figure}

For the question of whether the planetary system model used or the mass of the host star (and thus the stellar density in the vicinity of a planetary system) has a stronger influence on the formation of high eccentricities, we plot in Fig.~\ref{fig:cumulative_hist_ecc} the cumulative distribution of eccentricities after $100\,\mathrm{Myr}$. We distinguish between the three planetary system models (independent of the host-star mass) and the host-star mass (independent of the used planetary system model). For those systems that orbit a $2.5\,\mathrm{M}_\odot$ host star, it can be clearly seen that the host star mass, and thus the position in the cluster, which in turn is related to the stellar density in the vicinity, plays a more important role in exciting planets to high eccentricities than the exact orbital configuration and multiplicity of the planetary system.

Many studies have invesigated the critical engulfment distance during the giant branch phases at different levels of detail and with different underlying theories \citep[][]{Mustill2012,Adams2013,Nordhaus2013,Villaver2014,Madappatt2016,Privitera2016,Ronco2020}. We use the critical engulfment distances along the asymptotic giant branch phases for Earth-mass planets from figs.~2--4 in \cite{Mustill2012} and calculate for how many planets the periastron distance ($r_p = (1-e)a$) would be below this limit. For the planets around a $1.5\,\mathrm{M}_\odot$ star, the critical distance is about $1.9\,\mathrm{au}$. About 5 per cent of our planets around such a star would be engulfed during the giant branch phases. The critical distance for $2.0$ and $2.5\,\mathrm{M}_\odot$ stars is $2.2$ and $2.3\,\mathrm{au}$, respectively. In these cases, 16 per cent and 15 per cent of the planets would be engulfed, respectively. Although the possibility of being engulfed by the host star mainly affects the innermost planet P1, it is not exclusive, since an external perturbation combined with internal planet-planet scattering can cause even the initially outermost planet to migrate to a small or highly eccentric orbit, causing its periastron distance to fall below the critical value. This can be seen in the bottom panels of Fig.~\ref{fig:a-e_space}, where we additionally plot the critical engulfment distance for each host-star mass.

\subsection{Inclination and Retrograde Orbits}

The first exoplanets thought to have a polar or retrograde orbit ($i \ge 90^\circ$) were HAT-P-7 b \citep[][]{Winn2009} and WASP-17 b \citep[][]{Anderson2010, Bayliss2010}. Since the number of confirmed retrograde planetary orbits is still small, the statistical abundance of these peculiar orbits is still uncertain. In addition to the expected clustering of prograde orbits, \cite{Albrecht2021} recently found a further clustering of polar orbits in a sample of 57 systems rather than a scattering over the entire range of possible obliquities.
Since all planets in our simulations are initially co-planar, most planetary orbits are still only slightly inclined at the end of the simulations.

\begin{figure}
    \centering
    \includegraphics[width = 0.49\textwidth,trim= 0.6cm 0.2cm 1.1cm 0.2cm,clip]{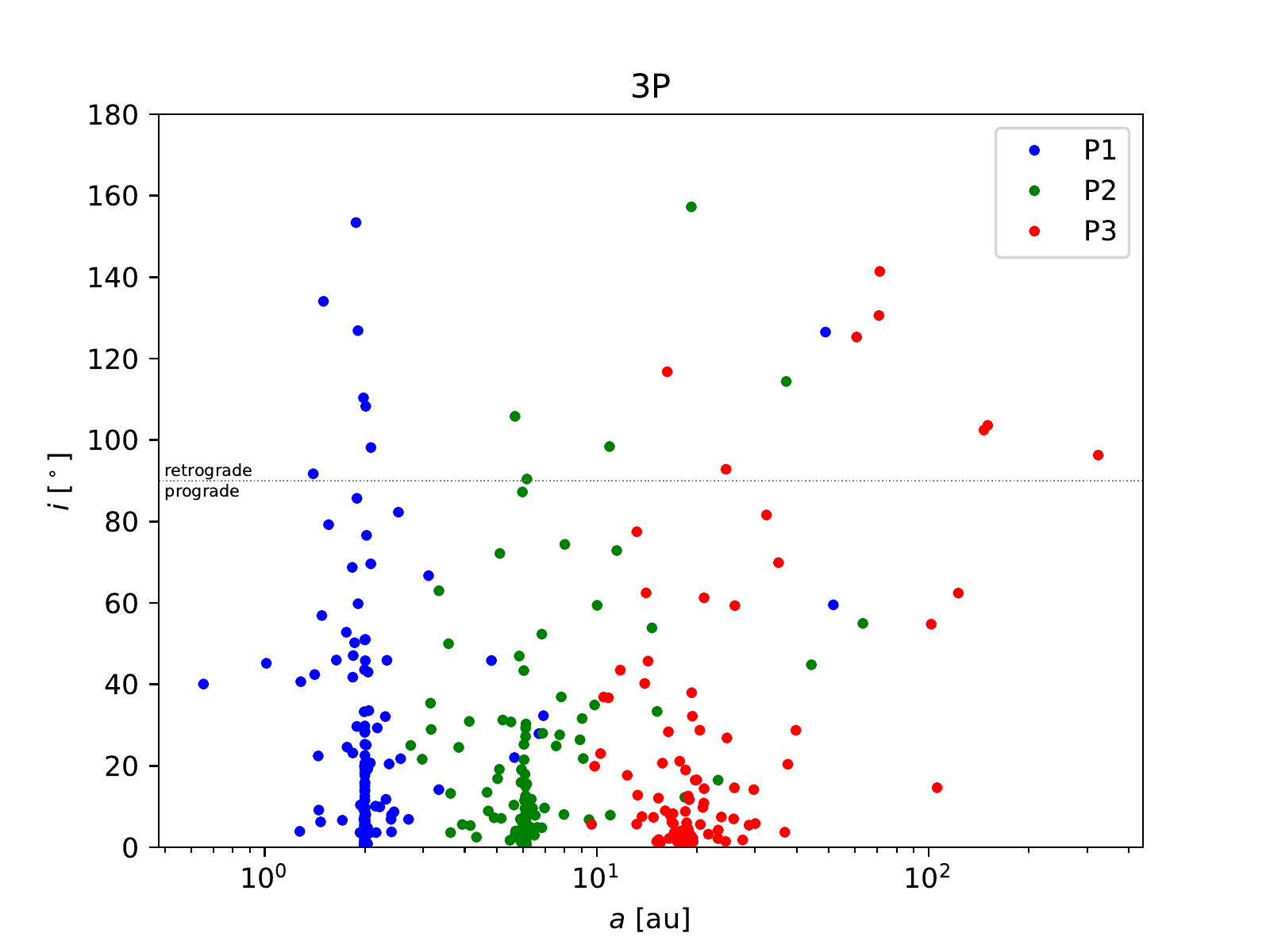}
    \includegraphics[width = 0.49\textwidth,trim= 0.6cm 0.2cm 1.1cm 0.2cm,clip]{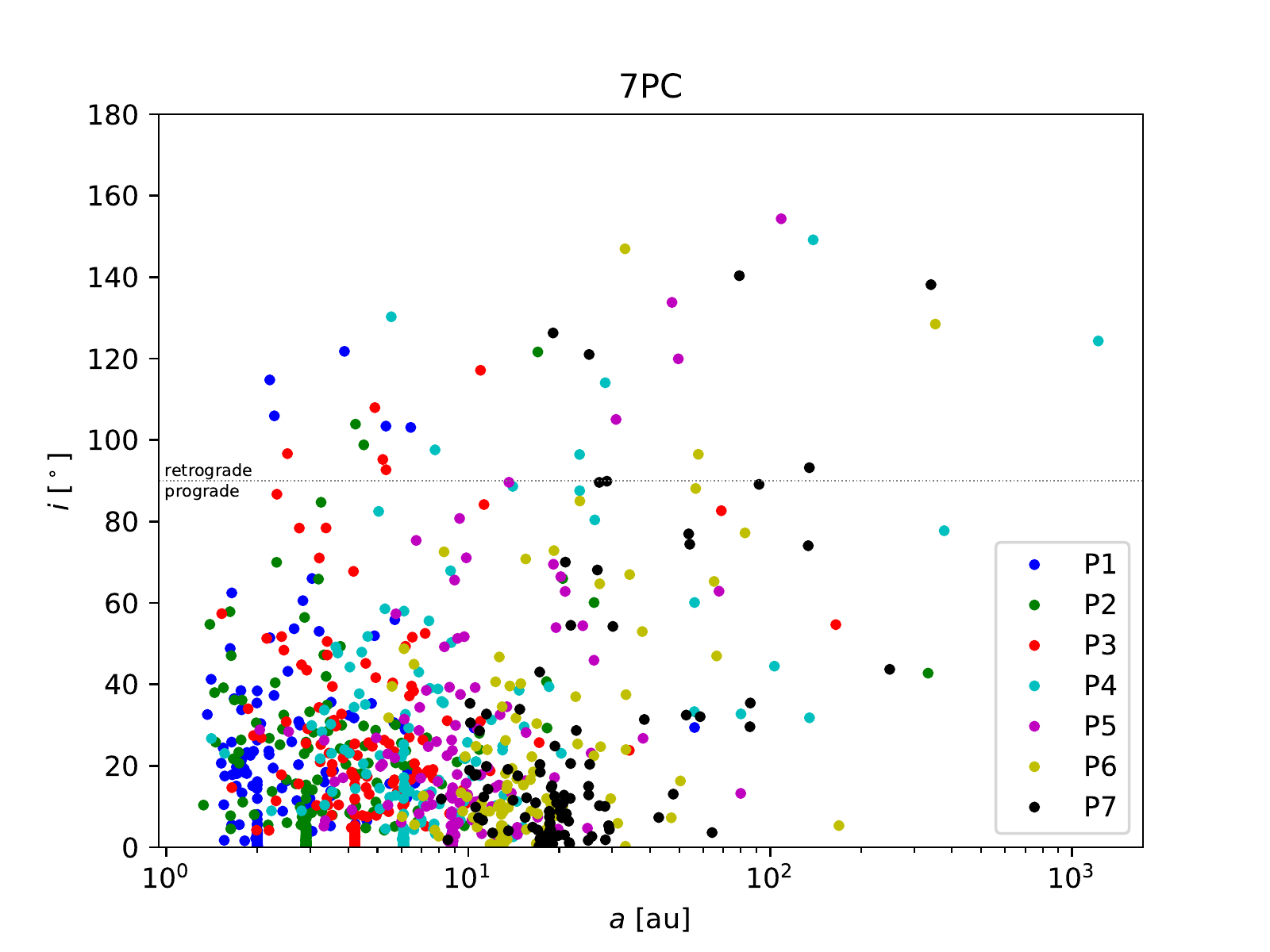}
    \includegraphics[width = 0.49\textwidth,trim= 0.6cm 0.2cm 1.1cm 0.2cm,clip]{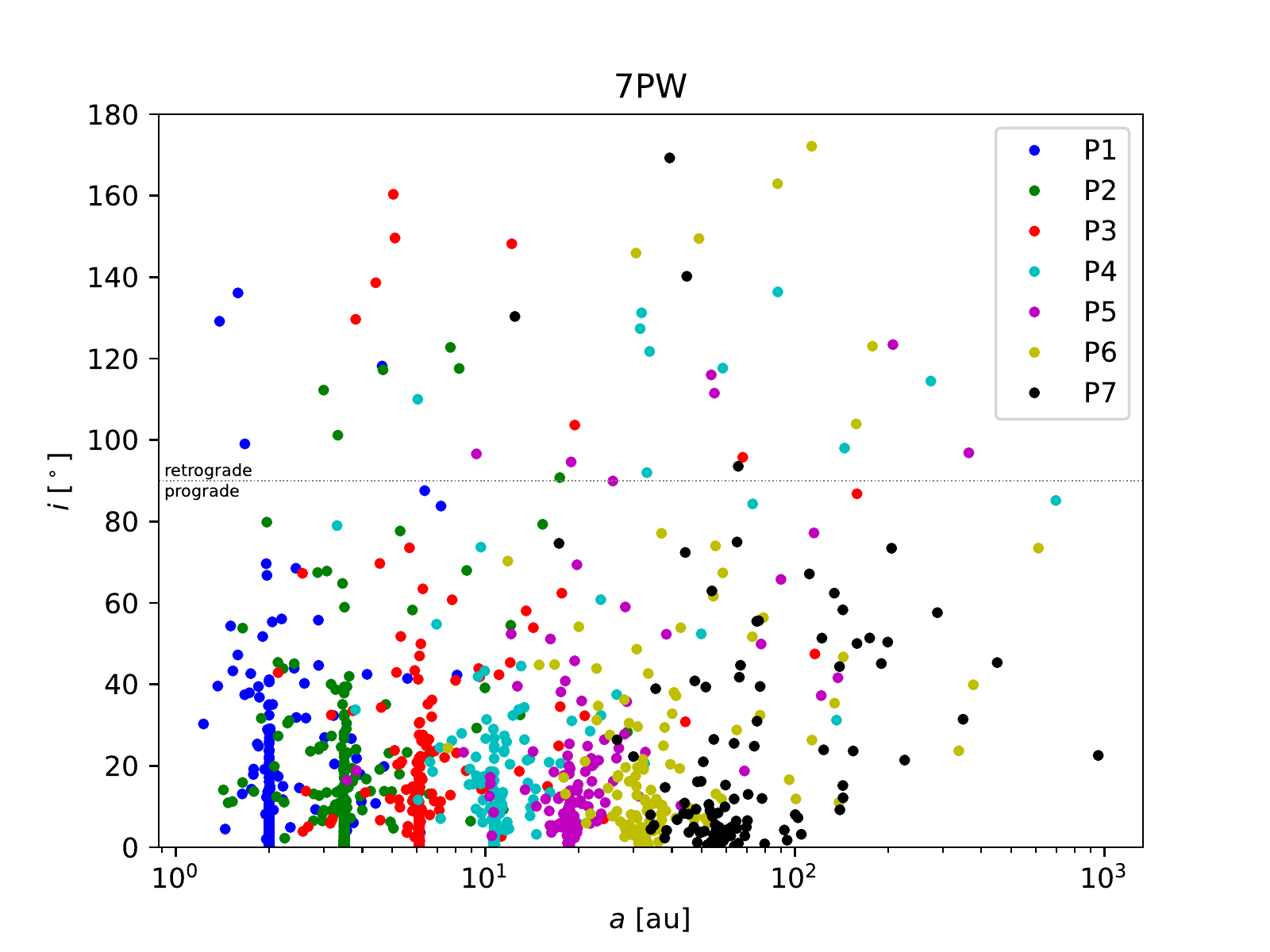}
    \caption{The $a$-$i$ space for the 3P, 7PC, and 7PW model. Planets above the dotted grey line are on a retrograde orbit.}
    \label{fig:a-i_space}
\end{figure}

The distribution of planetary orbits in the $a$-$i$ space for all three planetary system models at the end of the simulation is shown in Fig.~\ref{fig:a-i_space}.

The averaged fraction of planets with a retrograde motion at the end of the simulation is 1.4 per cent, with values ranging from 0.5--2.2 per cent for the different system models (the fraction of retrograde orbits for each scenario used in this work are listed in Table~\ref{tab:fraction_orbits}). This is somewhat higher but still in good agreement with the values in \cite{Stock2020}, where we used variations of the Solar System around Sun-like host stars and found, depending on the initial planet configuration and the star cluster size, 0.1--1.6 per cent of all planets to be on a retrograde orbit after the same simulation time of $100\,\mathrm{Myr}$. The subtle differences can be explained by the higher multiplicity in the 7PC and 7PW models as well as by the generally larger host star masses used in this study.
This agreement, and the circumstance that we cover a wide range of possible planetary systems in this work and in \cite{Stock2020}, leads us to the rough estimate that in open star clusters similar to the one simulated in this work and those simulated in \cite{Stock2020}, about 1--2 per cent of all planets could be on stable retrograde orbits.

The number of planets that flip to a retrograde orbit for at least one integration step at some time during the simulation is significantly larger and gives an indication that unstable retrograde orbits are not uncommon in environments with frequent external gravitational perturbation. ``Unstable retrograde orbit'' in this context means that the planet does not remain permanently on a retrograde orbit, either because it changes back to a prograde orbit or because it is ejected from the planetary system at a later time. In 33 per cent of all systems we find at least one planet which flips to a retrograde orbit for at least one (stored) time step of 1,000 years during the simulation. This fraction of systems is generally lowest for the 3P model and $1.5\,\mathrm{M}_\odot$ stars, and highest for the 7PW model and $2.5\,\mathrm{M}_\odot$ stars.

\begin{figure}
    \centering
    \includegraphics[width= 0.47\textwidth,trim= 0.6cm 0.2cm 1.25cm 0.75cm,clip]{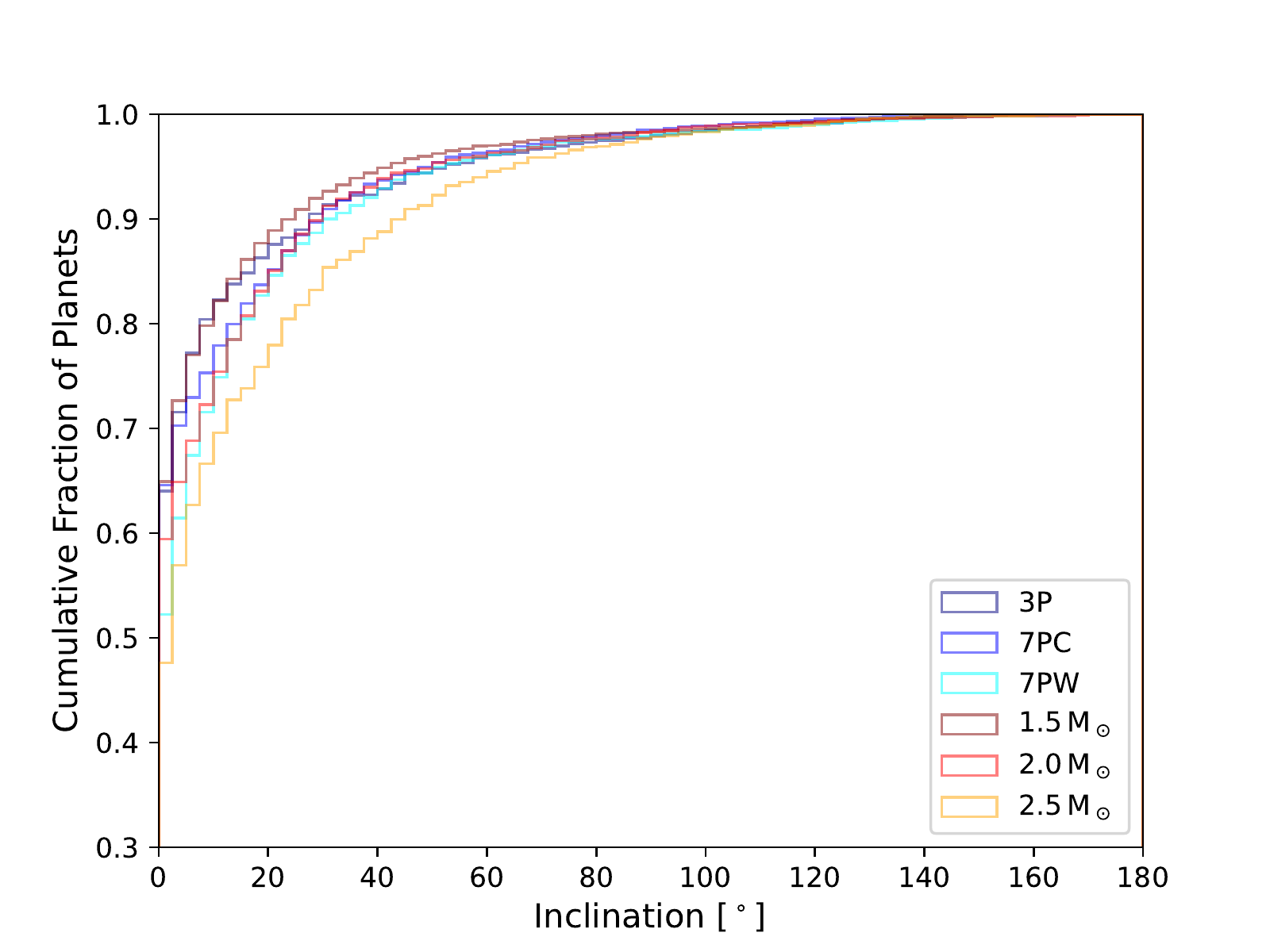}
    \caption{Cumulative, normalized histogram showing the distribution of inclinations sorted by planetary model (bluish colors) and host star mass (reddish colors). The bin size is $2.5^\circ$.}
    \label{fig:cumulative_hist_inc}
\end{figure}

Since inclined orbits, just like eccentric orbits, from through angular momentum exchange, and close stellar encounters are particularly good at introducing an angular-momentum deficit into the planetary system, it is especially the systems around $2.5\,\mathrm{M}_\odot$ stars which have inclined orbits, as can be seen in Fig.~\ref{fig:cumulative_hist_inc}. Again, the mass of the host star and thus the frequency and strength of the encounters with other cluster members is more important than the exact planetary configuration for the formation of inclined planetary orbits.

\subsection{Mean-Motion Resonances}

Resonances in planetary systems can be a source for WD pollution, because the orbits of asteroids can increase in eccentricity due to planets near a secular or mean-motion resonance (MMR) \citep[][]{Debes2012,Smallwood2018,Smallwood2021,Antoniadou2019,Veras2021b}. Finding MMRs in simulations is very challenging in view of the large number of simulated planetary systems, the long integration time, but especially because of the often chaotic dynamical evolutions of the planetary systems due to the steady external perturbation from the cluster. This type of dynamical evolution can lead to transitory resonances which may endure over just a handful of output timesteps, or even within two consecutive outputs of the simulations. 

We use the FAIR method of \cite{Forgacs2018} which allows the fast identification of MMRs between planets without any prior knowledge about the MMR to be searched. The method is based on plotting the difference of the mean orbital longitudes for two planets ($\lambda^\prime - \lambda$, if $a < a^\prime$) against the mean anomaly $M$ of the inner planet. The mean longitude is defined as $\lambda = M + \varpi  = M + \omega + \Omega$, where $\varpi$ is the longitude of the periastron and $\omega$ the argument of periastron. When the planets have a mean-motion ratio of $n/n^\prime = (p+q)/p$ throughout the period under consideration, there will be $q$ centres on the x-axis and $p+q$ centres on the y-axis, which in turn means that only the number of crossings of the stripes present in the plot with the horizontal and vertical axes must be counted to obtain $q$ and $p+q$, respectively. Subsequently, the necessary criteria for the presence of an MMR, the period ratios and the libration of the resonance variables

\begin{equation}
\theta_1 = (p+q)\lambda^\prime - p\lambda -q \varpi,
\end{equation}
\begin{equation}
\theta_2 = (p+q)\lambda^\prime - p\lambda -q \varpi^\prime,
\end{equation}
around a mean value, need to be tested. This value is not necessarily always $0^{\circ}$ or $180^{\circ}$.

In principle, we find MMR in our simulations at arbitrary times (except at the beginning), but they only rarely survive for longer times (i.e. several million years) due to the constant gravitational perturbation of the neighbouring stars.
Only with increasing simulation duration and the expansion of the host star cluster, when the frequency and strength of the encounters with neighbouring stars decrease, can the planets actually remain in MMR for several million years.
In particular, we therefore investigate how many and which MMRs are found in the last 1 million years of our simulations that are stable until the end of the simulation at $t = 100\,\mathrm{Myr}$, since these can also persist once the cluster has completely dissolved. If we consider only planetary pairs that were direct neighbours at the beginning of the simulation, we find six planetary pairs that are in stable MMR at the end of the simulation and list them in Table~\ref{tab:MMR}.

\begin{figure}
    \centering
    \includegraphics[width=0.5\textwidth,trim= 1.4cm 0.1cm 1.3cm 0.5cm,clip]{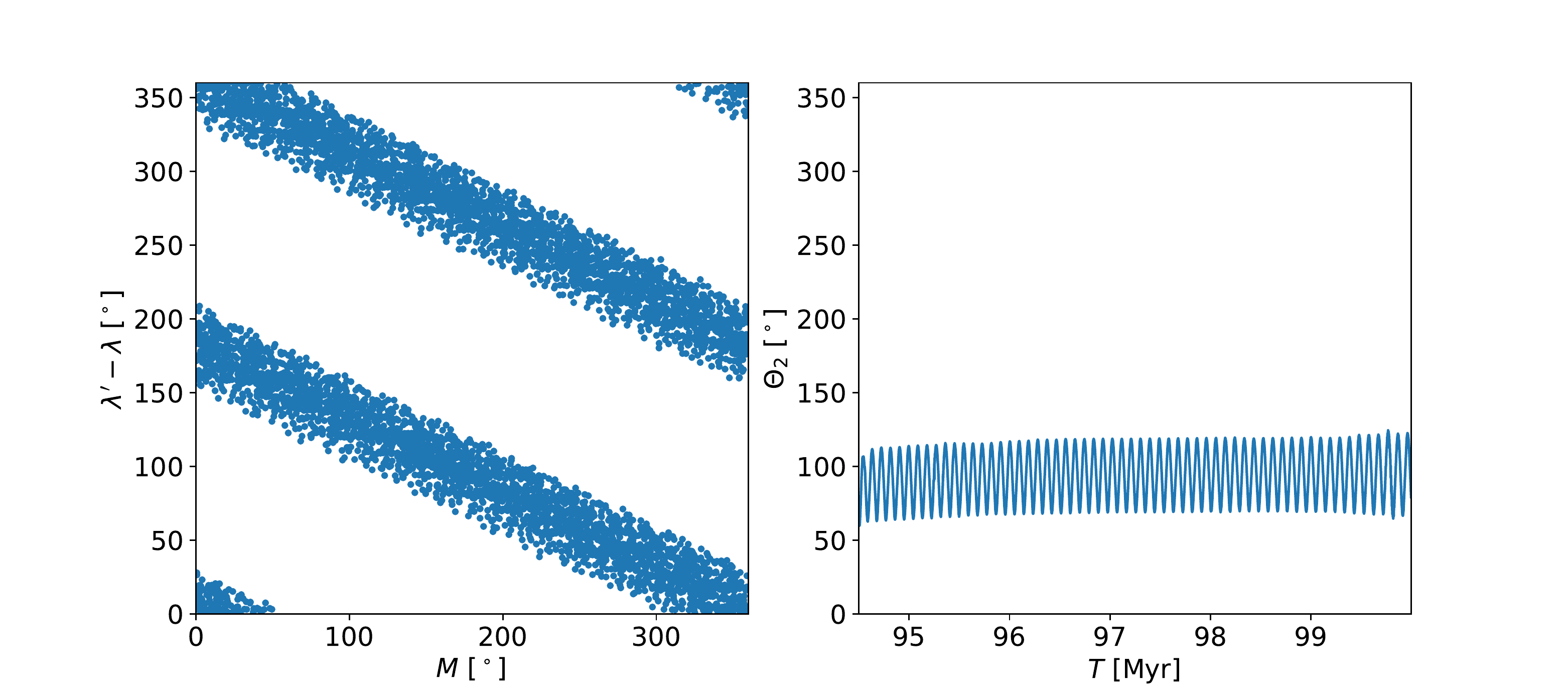}
    \caption{2:1 MMR in planetary system 122 (7PC model, $1.5\,\mathrm{M_\odot}$ host star) between planet 6 and 7 for $t=94.5$--$100\,\mathrm{Myr}$.}
    \label{fig:MMR_p_sys_122_7PC_1_5MSun}
\end{figure}

\begin{figure}
    \centering
    \includegraphics[width=0.49\textwidth,trim= 0.8cm 0.3cm 0.5cm 0.5cm,clip]{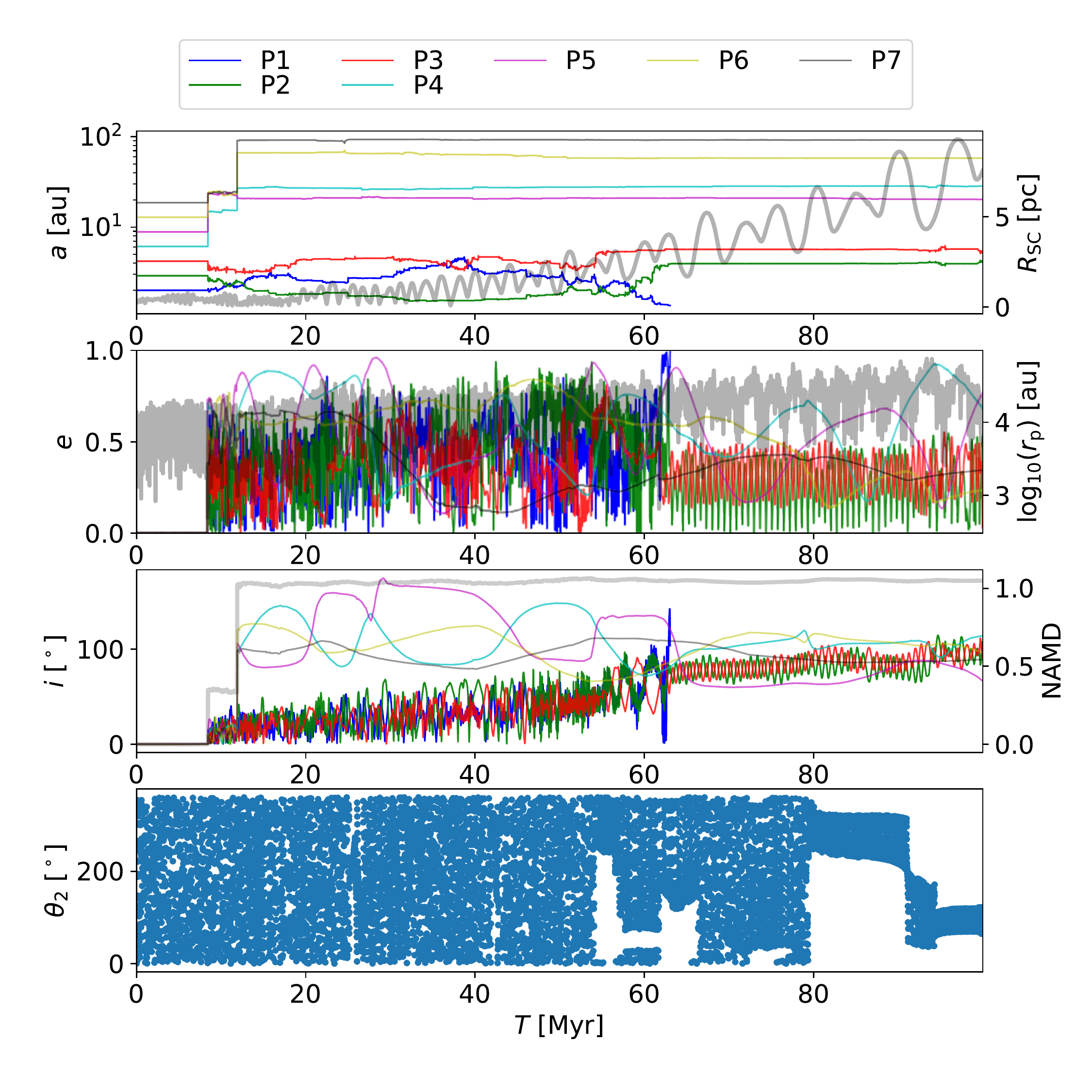}
    \caption{Orbital elements $a$, $e$ and $i$ for planetary system 122 (7PC model, $1.5\,\mathrm{M_\odot}$ host star) and the resonance angle $\theta_2$ with $p=q=1$ (for planet 6 and 7) as a function of the simulation time. The thicker grey lines in the background represent the distance to the cluster centre (top panel), the distance to the closest stellar perturber (top middle panel) and the normalized angular momentum deficit (bottom middle panel), defined as $\mathrm{NAMD} = \mathrm{AMD}/(\sum_k m_k \sqrt{\mathrm{G}m_\star a_k})$ \citep[see][and references therein]{Turrini2020}.}
    \label{fig:elements_p_sys_122_7PC_1_5MSun}
\end{figure}

For the 2:1 MMR in system 122 from the 7PC model with a $1.5\,\mathrm{M}_\odot$, we show the FAIR plot for the time range $t=99$--$100\,\mathrm{Myr}$ in Fig.~\ref{fig:MMR_p_sys_122_7PC_1_5MSun} and, in addition, also plot the orbital elements as well as the resonance angle of the 2:1 MMR as a function of the total simulation time in Fig.~\ref{fig:elements_p_sys_122_7PC_1_5MSun}. This makes it possible to see whether a system was excited into this resonance by chance only at the end of the simulation or whether the planets were already close to resonance for a long time.

As can be seen in Fig.~\ref{fig:elements_p_sys_122_7PC_1_5MSun}, planets 6 and 7 both migrate outwards to eccentric crossing orbits with $a\sim 23$--$24\,\mathrm{au}$ as a result of a close encounter ($r_p < 231\,\mathrm{au}$) at $t=8.4\,\mathrm{Myr}$ with a $0.2\,\mathrm{M}_\odot$ star. After another encounter at time $t=11.8\,\mathrm{Myr}$, both planets are thrown to orbits of $a=66\,\mathrm{au}$ and $a=91\,\mathrm{au}$, and thus happen to be near 2:1 MMR. Due to several further, but weak perturbations, they first enter 2:1 MMR at $t=54.2\,\mathrm{Myr}$, which does not completely resolve until $t=66.2\,\mathrm{Myr}$ (however, the libration amplitude is very large in the interim). Due to subsequent weak perturbations, the planets re-enter 2:1 MMR at $t=80\,\mathrm{Myr}$. The angle around which the resonance angle librates changes during the remaining simulation time, but in principle the planets remain in 2:1 MMR until the end of the simulation. That both planets are in stable resonance over the last one million years can be seen in Fig.~\ref{fig:MMR_p_sys_122_7PC_1_5MSun}.

Planet-planet scattering as a cause of MMRs is often disregarded. Yet they can be particularly responsible for the higher-order resonances \citep[][]{Raymond2008}, as these cannot arise so easily through migration \citep[][]{Papaloizou2005,Tadeu2015,Xu2018}. In our simulations, however, star-planet interactions seem to be more important than planet-planet interactions. In addition to four first-order resonances, we find a third-order resonance and a seventh-order resonance, and plot the resonances (Figs.~\ref{fig:MMR_p_sys_177_3P_1_5MSun}, \ref{fig:MMR_p_sys_22_7PW_2MSun}, \ref{fig:MMR_p_sys_45_7PW_2MSun}, \ref{fig:MMR_p_sys_38_7PW_2_5MSun}, \ref{fig:MMR_p_sys_192_7PW_2_5MSun} in the appendix) and the overall dynamical evolution of the systems including perturber information (Figs.~\ref{fig:elements_p_sys_177_3P_1_5MSun}, \ref{fig:elements_p_sys_22_7PW_2MSun}, \ref{fig:elements_p_sys_45_7PW_2MSun}, \ref{fig:elements_p_sys_38_7PW_2_5MSun}, \ref{fig:elements_p_sys_192_7PW_2_5MSun} in the appendix) for each of these systems. However, some librating resonance angles show a long-term trend that cannot be resolved in time. Whether the systems are in actual --- and stable --- resonance cannot be said with certainty in these cases. In all systems, one or several encounters with neighbouring stars lead to a migration of the planetary pairs near a certain MMR. Further weaker stellar perturbations, usually millions of years later, drive the planetary pair into actual resonance. Only in the case of system 192 from the 7PW model around a $2.5\,\mathrm{M}\odot$ star (see Fig.~\ref{fig:elements_p_sys_192_7PW_2_5MSun} in the appendix) a resonance is created directly by the first strong stellar perturbation, but this resonance is repeatedly perturbed by stellar neighbours at later times without breaking it completely.

Our results are only partially comparable to \cite{Raymond2008}, since the instability in our simulations has an external rather than an internal origin. In most cases, the external perturbation completely outweighs internal effects. In cases where planet-planet scattering may play an additional role in the origin of the resonance, the effects of external and internal perturbation cannot be separated clearly enough. However, our simulations confirm that most of the resonances that arise are low-order and that higher-order resonances can also arise in a few cases. Furthermore, an important difference from the simulations in \cite{Raymond2008} is the mass distribution within the planetary system. While we exclusively simulate planets of equal masses, \cite{Raymond2008} also use mixed systems, in which the interplay of large and small planet plays an important role in the formation of resonances due to planet-planet scattering.

\begin{table}
    \centering
    \begin{tabular}{cc|cccc}
         \hline
         Host Star & Model & System ID & Planet Pair & MMR & $t\,$[Myr]  \\
         \hline
          $1.5\,\mathrm{M}\odot$ & 3P & 177 & 2/3 & 10:3 & 97.0--100 \\
          $1.5\,\mathrm{M}\odot$ & 7PC & 122 & 6/7 & 2:1 & 95.0--100\\
          $2.0\,\mathrm{M}\odot$ & 7PW & 22 & 5/6 & 3:2 & 99.0--100 \\
          $2.0\,\mathrm{M}\odot$ & 7PW & 45 & 5/6 & 3:2 & 90.0--100\\
          $2.5\,\mathrm{M}\odot$ & 7PW & 38 & 4/5 & 5:2 & 97.0--100 \\
          $2.5\,\mathrm{M}\odot$ & 7PW & 192 & 3/4 & 4:3 & 98.5--100 \\
         \hline
    \end{tabular}
    \caption{Stable MMRs after the end of the simulation.}
    \label{tab:MMR}
\end{table}

\subsection{Long-term Stability}

The focus of this investigation is to perform stellar cluster simulations including multi-planet systems, and to present the results as initial conditions which may be used for the simulation community. Nevertheless, we can make some preliminary rapid judgements about the stability of these systems along the main sequence by taking advantage of machine learning.

In order to estimate the long-term stability of each planetary system (with the host star mass remaining constant), we perform a \texttt{SPOCK} test \citep[][]{Tamayo2020b} for each system. \texttt{SPOCK} is able to predict the long-term stability of compact multi-planet systems by using the statistics from machine learning training datasets in a fraction of the time compared to actual integration \citep[see][for a detailed description of the training and the model]{Tamayo2020b}. Since \texttt{SPOCK} requires at least three planets in the system, for those systems where planets have been ejected, we replace the missing planets with massless particles at wide orbits ($> 100\,\mathrm{au}$) before performing the \texttt{SPOCK} test. We give the likelihood for the long-term stability of our planetary systems as additional feature in our result tables (see online or extracts in Tables \ref{tab:results_1_5MSun_3P} and \ref{tab:results_2_5MSun_7PW} in the appendix) for the different planet system models.

\section{Conclusions}

We have presented the results from a total of 1224 simulations of planetary systems embedded in a cluster environment, differing in the planetary system model used and in the mass of the host star. We have aimed to publish a comprehensive dataset that contains those planetary systems typically responsible for WD pollution which, at the same time, have the dynamical imprint of a typical birth star cluster. This data set can now be used for further numerical integration beyond the main sequence to the WD phase. 

The three different planetary system models should represent the extreme cases of possible planetary systems regarding their multiplicity and orbital spacing. For this reason we simulate one model with only three planets (3P model), and two models with seven planets each in the system, which differ in their compactness (7PC and 7PW model). As host stars we have chosen those stars from our 8000 star cluster which are most similar to the masses of $1.5\,\mathrm{M}_\odot$, $2.0\,\mathrm{M}_\odot$ and $2.5\,\mathrm{M}_\odot$. All planetary systems are integrated to $t=100\,\mathrm{Myr}$. By that time the star cluster has expanded considerably and perturbations from neighbouring stars are very rare.
 
In our simulations, it was not only the number of planets or the compactness of the system that played a role in the average survival rate, but above all the semimajor axes of the outermost planets. The 3P model has an average survival fraction of 76 per cent, the 7PC model has 74 per cent and the 7PW model, with the widest orbits, has only 71 per cent. While the innermost planet has the highest survival probability in the 3P model, it is the third planet in the 7PC model and the second innermost planet in the 7PW model. The spread in survivability has been particularly small in the 7PC model, which is why we conclude that the compactness of the system and thus enhanced internal effects such as planet-planet scattering almost equalizes the planets' probability to survive the star cluster phase. 

Given our initial conditions, we found that about 5 per cent of planets around $1.5\,\mathrm{M}_\odot$ stars, roughly 16 per cent of the planets around $2.0\,\mathrm{M}_\odot$, and approximately 15 per cent of the planets around $2.5\,\mathrm{M}_\odot$ stars would be swallowed by the eventual asymptotic giant branch star's envelope because the planets' periastron distances would be below the critical engulfment distance.

The excitation in eccentricity and inclination correlates with the number of planets in the system and the initial semimajor axis of the outermost planet. In particular it also correlates with the stellar density in the vicinity of the host star, which tends to be larger for higher mass stars due to the effect of mass segregation. On average, 1.4 per cent of all planets are on a retrograde orbit at the end of the simulation, which is, due to the higher host star masses and the higher multiplicity in the 7PC and 7PW model, somewhat higher but still in good agreement with the results from \cite{Stock2020}.

Eccentric planets in the super-Earth mass regime are thought to be particularly efficient drivers for WD pollution over a wide range of cooling ages \citep[][]{Frewen2014, Mustill2018}. 30 per cent of the planets in our simulations attained an eccentricity of $e>0.1$ and 25 per cent have $e>0.17$ after $100\,\mathrm{Myr}$, showing that the birth environment of planetary systems can produce a sufficient distribution in eccentricity to help generate the architectures suitable for dynamical delivery of pollutants to WDs. Even if subsequent increases in eccentricity due to mutual perturbations and/or along the giant branch phases due to stellar mass loss alone \citep{Veras2011} are negligible, the planets' primordial eccentricities will persist into the WD phase.

Furthermore, we find planetary pairs in several planetary systems that are in resonance at the end of the simulation. These systems may also play a role in WD pollution, since asteroids near these resonances may be driven to eccentric orbits and subsequently be tidally disrupted by the WD \citep[][]{Smallwood2018,Smallwood2021}.

\section*{Acknowledgements}

We thank the reviewer for helpful comments which have improved the manuscript. We would also like to thank Dr. Martin Schlecker for the valuable discussions on the initial conditions of our simulations. KS and RS acknowledge the support of the DFG priority program SPP 1992 ``Exploring the Diversity of Extrasolar Planets'' (SP 345/20-1 and SP 345/22-1). DV  gratefully acknowledges the support of the STFC via an Ernest Rutherford Fellowship (grant ST/P003850/1). The simulations were carried out partly on the GPU accelerated cluster ``kepler'', funded by Volkswagen Foundation grants 84678/84680. The authors acknowledge support by the High Performance and Cloud Computing Group at the Zentrum für Datenverarbeitung of the University of T\"ubingen, the state of Baden-W\"urttemberg through bwHPC and the German Research Foundation (DFG) through grant no INST 37/935-1 FUGG.

\section*{Data Availability}
The results of the planetary simulations are available as supplementary material in the online version of this paper. The data underlying this article are available on Zenodo at \href{https://doi.org/10.5281/zenodo.5883613}{doi:10.5281/zenodo.5883613} in time steps of 2000 years. The data underlying this article will be shared in full temporal resolution (time steps of 1000 years) on reasonable request to the corresponding author.



\bibliographystyle{mnras}




\appendix



\newpage
\section{Mean-Motion Resonances at the End of the Simulations}

\begin{figure}
    \centering
    \includegraphics[width=0.5\textwidth,trim= 1.4cm 0.1cm 1.3cm 0.5cm,clip]{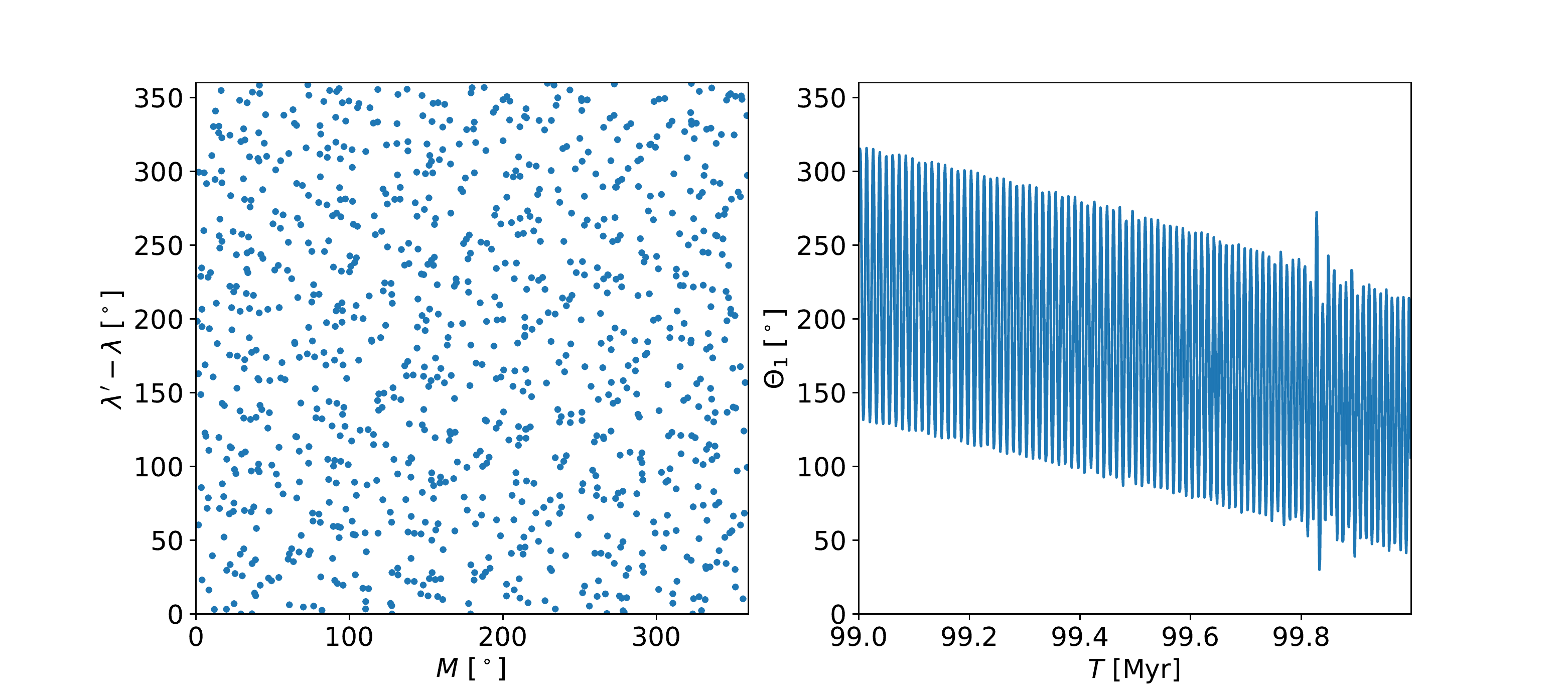}
    \caption{10:3 MMR between planet 2 and 3 in planetary system 177 (3P model, $1.5\,\mathrm{M_\odot}$ host star) between $99$--$100\,\mathrm{Myr}$.}
    \label{fig:MMR_p_sys_177_3P_1_5MSun}
\end{figure}

\begin{figure}
    \centering
    \includegraphics[width=0.5\textwidth]{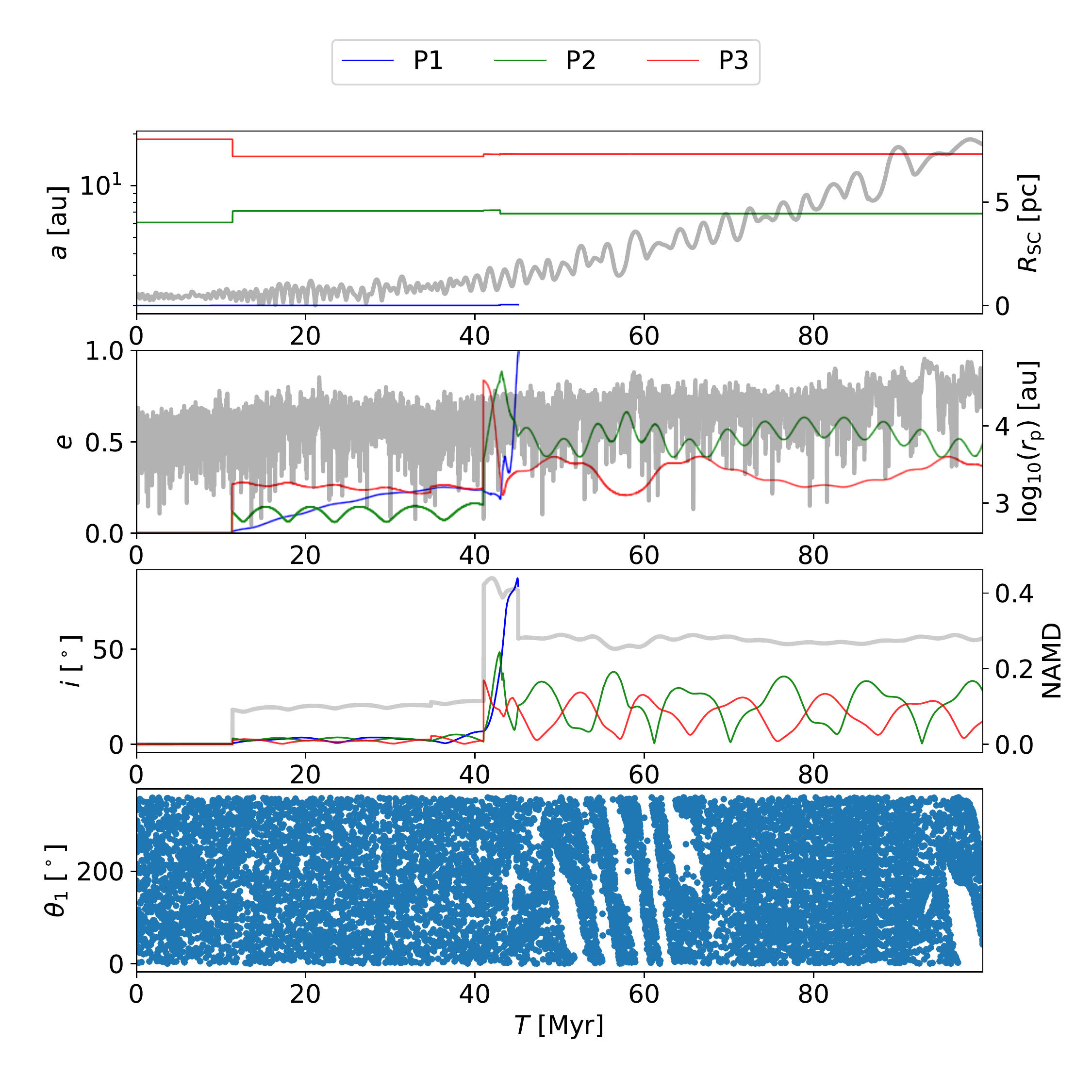}
    \caption{As Fig.~\ref{fig:elements_p_sys_122_7PC_1_5MSun} but for planetary system 177 (3P model, $1.5\,\mathrm{M_\odot}$ host star). The resonance angle is shown for $p=3$, $q=7$ and planet pair 2/3.}
    \label{fig:elements_p_sys_177_3P_1_5MSun}
\end{figure}

\begin{figure}
    \centering
    \includegraphics[width=0.5\textwidth,trim= 1.4cm 0.1cm 1.3cm 0.5cm,clip]{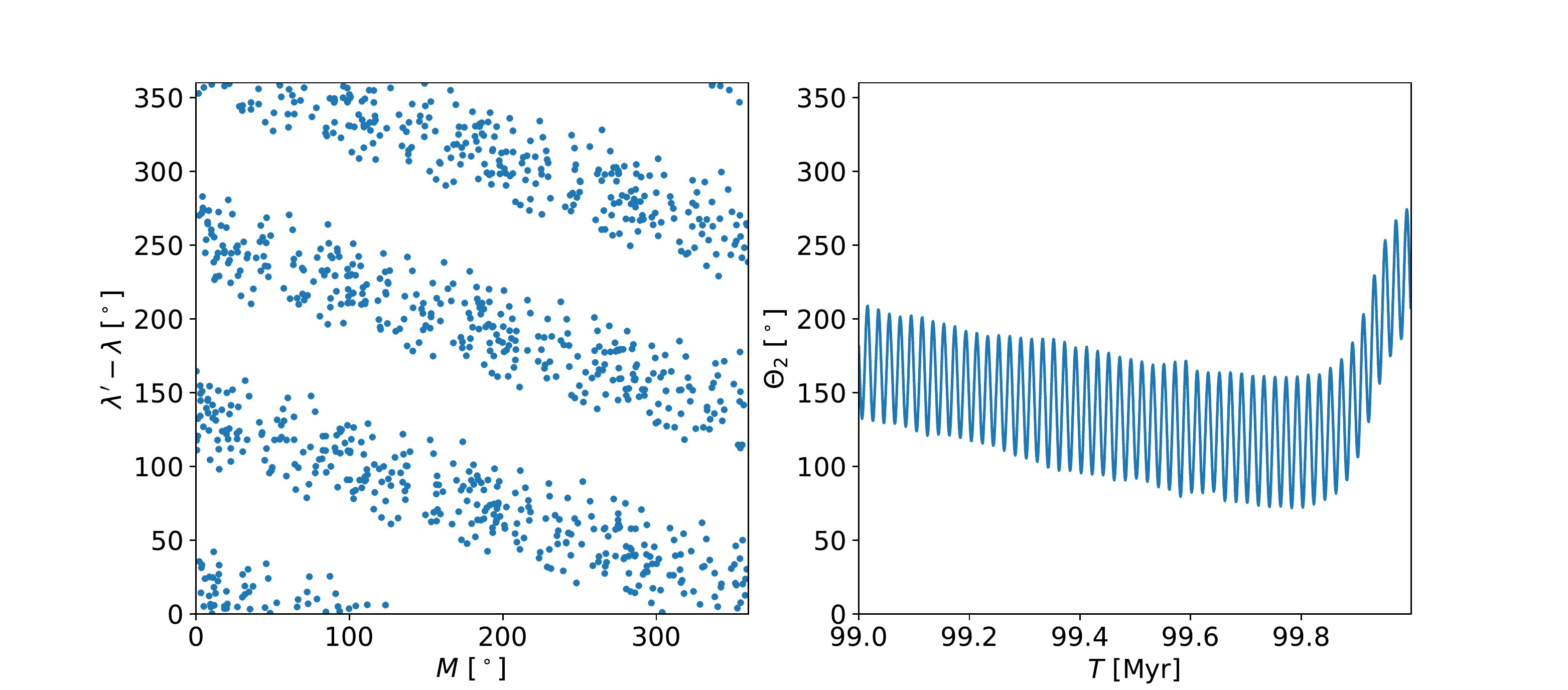}
    \caption{3:2 MMR  between planet 5 and 6 in planetary system 22 (7PW model, $2.0\,\mathrm{M_\odot}$ host star) between $99$--$100\,\mathrm{Myr}$.}
    \label{fig:MMR_p_sys_22_7PW_2MSun}
\end{figure}

\begin{figure}
    \centering
    \includegraphics[width=0.5\textwidth]{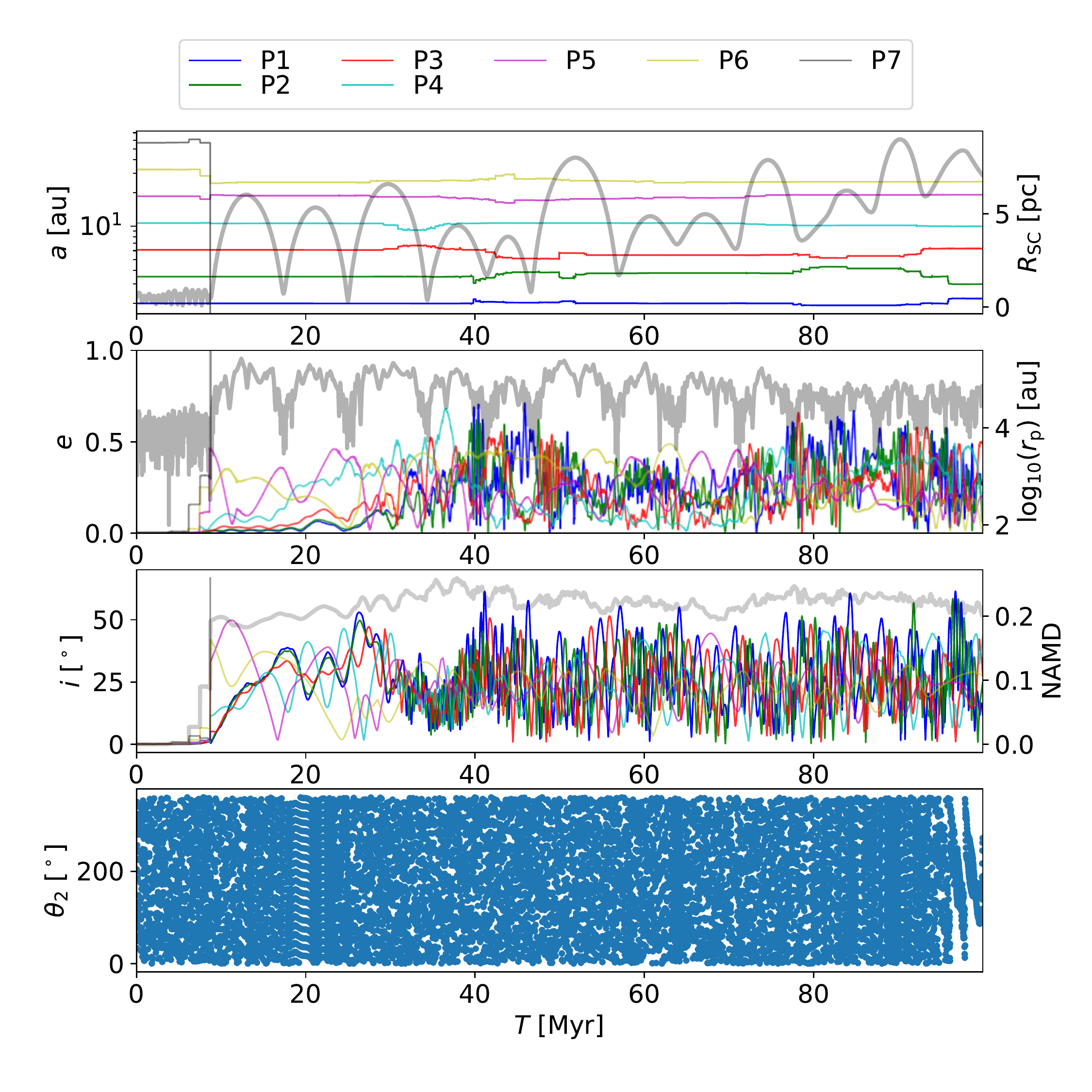}
    \caption{As Fig.~\ref{fig:elements_p_sys_122_7PC_1_5MSun} but for planetary system 22 (7PW model, $2.0\,\mathrm{M_\odot}$ host star). The resonance angle is shown for $p=2$, $q=1$ and planet pair 5/6}
    \label{fig:elements_p_sys_22_7PW_2MSun}
\end{figure}

\begin{figure}
    \centering
    \includegraphics[width=0.5\textwidth,trim= 1.4cm 0.1cm 1.3cm 0.5cm,clip]{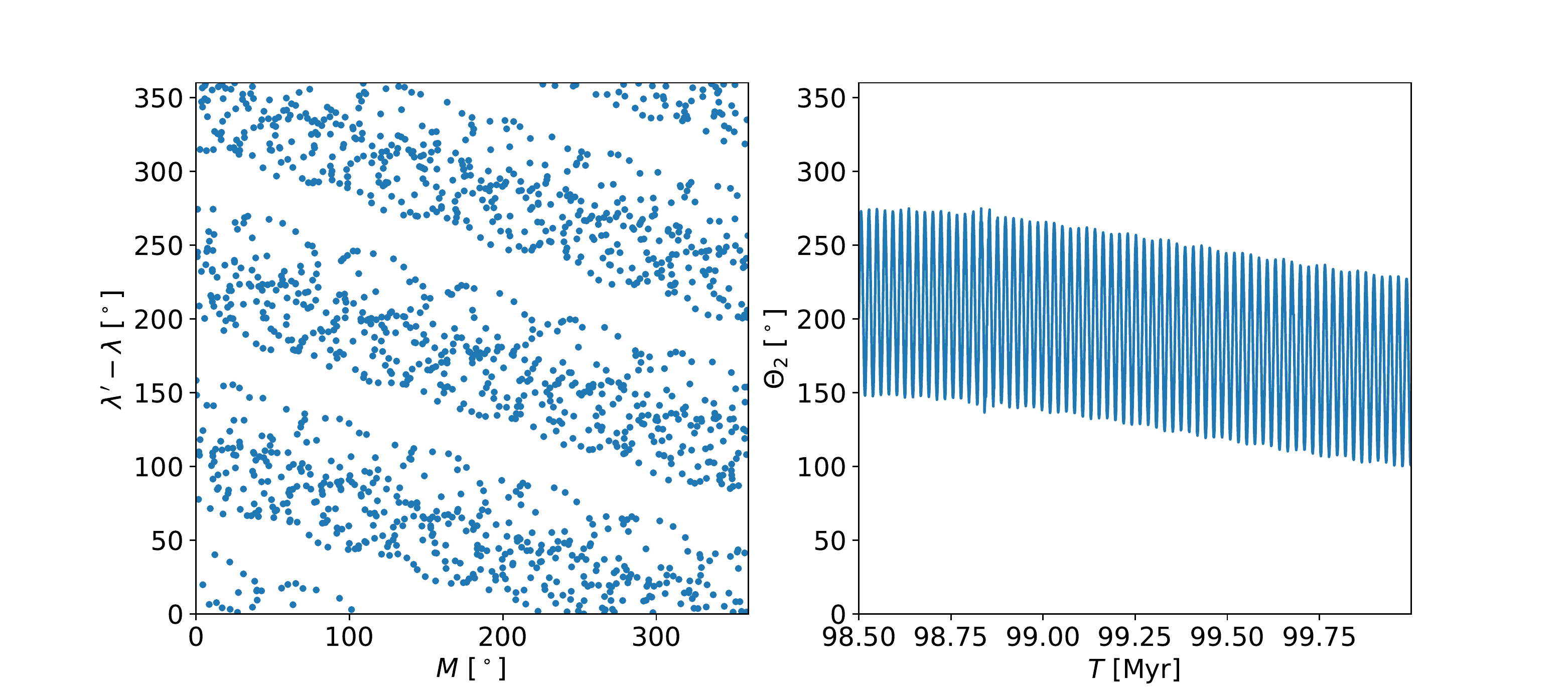}
    \caption{3:2 MMR between planet 5 and 6 in planetary system 45 (7PW model, $2.0\,\mathrm{M_\odot}$ host star) between $98.5$--$100\,\mathrm{Myr}$.}
    \label{fig:MMR_p_sys_45_7PW_2MSun}
\end{figure}

\begin{figure}
    \centering
    \includegraphics[width=0.5\textwidth]{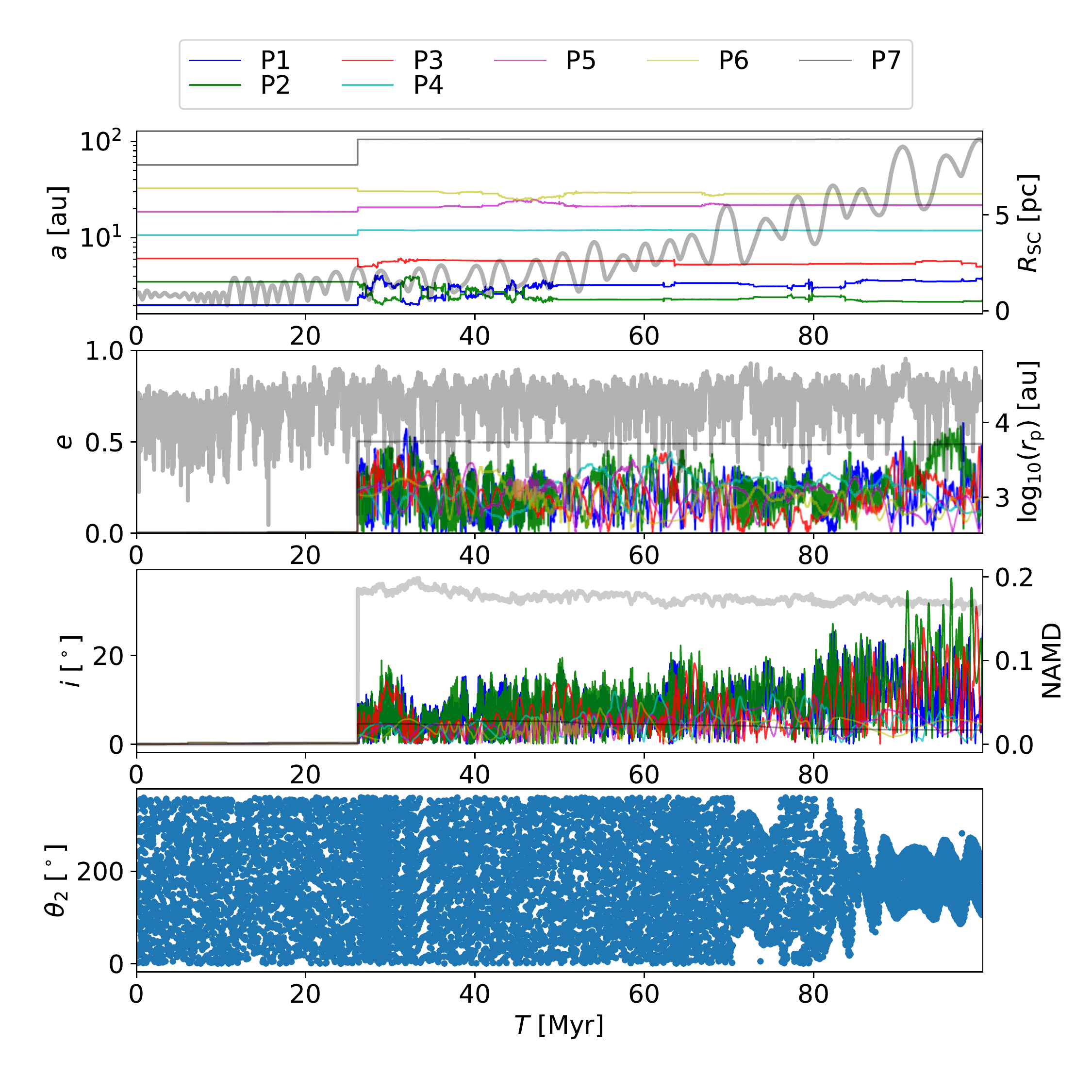}
    \caption{As Fig.~\ref{fig:elements_p_sys_122_7PC_1_5MSun} but for planetary system 45 (7PW model, $2.0\,\mathrm{M_\odot}$ host star). The resonance angle is shown for $p=2$, $q=1$ and planet pair 5/6.}
    \label{fig:elements_p_sys_45_7PW_2MSun}
\end{figure}

\begin{figure}
    \centering
    \includegraphics[width=0.5\textwidth,trim= 1.4cm 0.1cm 1.3cm 0.5cm,clip]{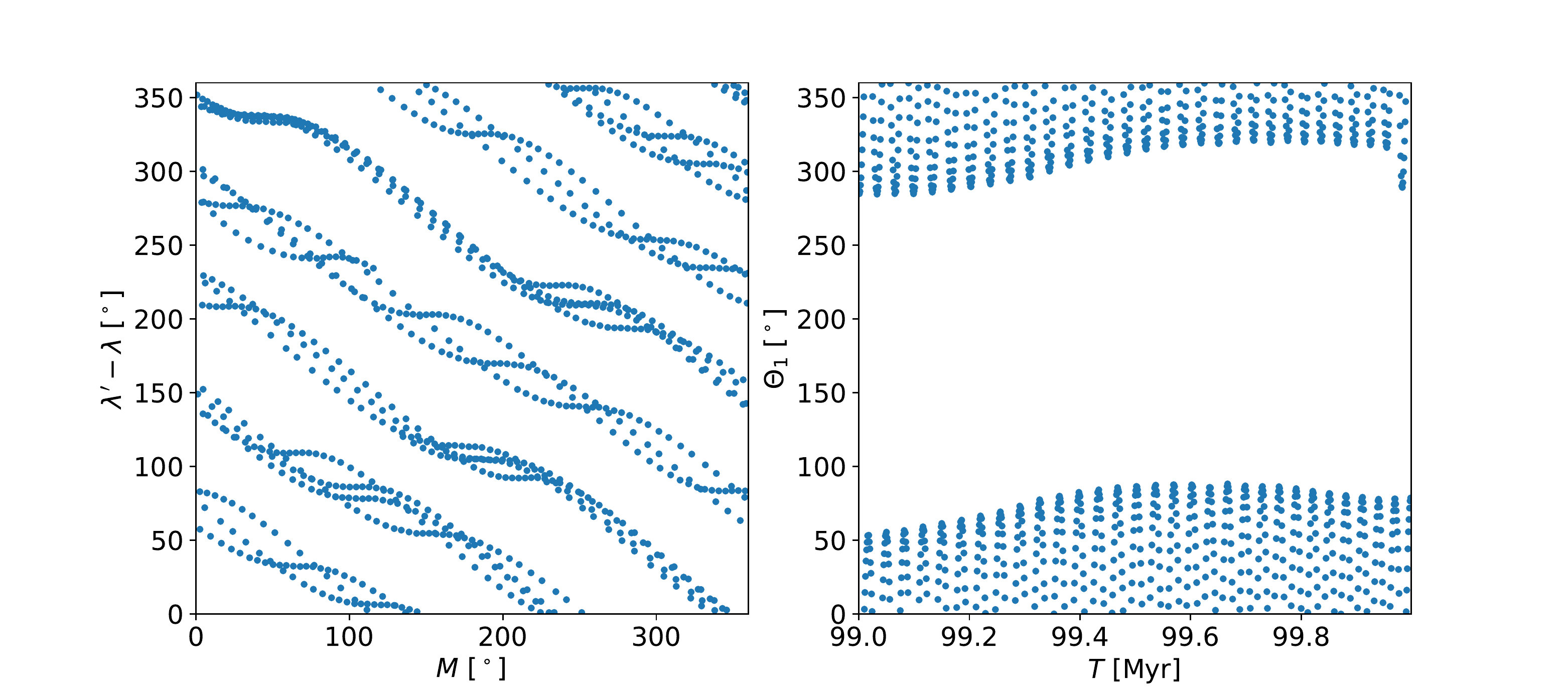}
    \caption{5:2 MMR between planet 4 and 5 in planetary system 38 (7PW model, $2.5\,\mathrm{M_\odot}$ host star) between $99$--$100\,\mathrm{Myr}$.}
    \label{fig:MMR_p_sys_38_7PW_2_5MSun}
\end{figure}

\begin{figure}
    \centering
    \includegraphics[width=0.5\textwidth]{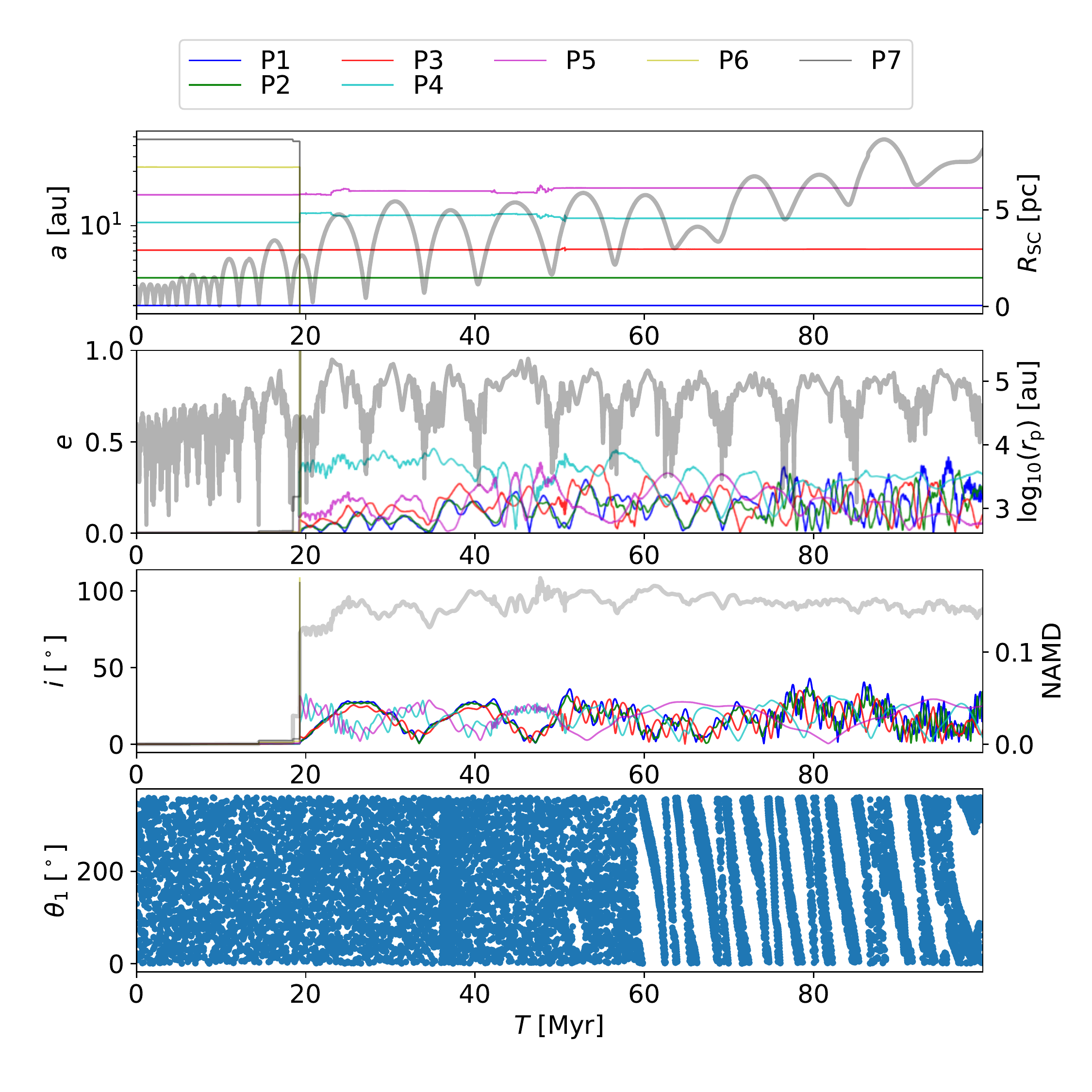}
    \caption{As Fig.~\ref{fig:elements_p_sys_122_7PC_1_5MSun} but for planetary system 38 (7PW model, $2.5\,\mathrm{M_\odot}$ host star). The resonance angle is shown for $p=2$, $q=3$ and planet pair 4/5.}
    \label{fig:elements_p_sys_38_7PW_2_5MSun}
\end{figure}

\begin{figure}
    \centering
    \includegraphics[width=0.5\textwidth,trim= 1.4cm 0.1cm 1.3cm 0.5cm,clip]{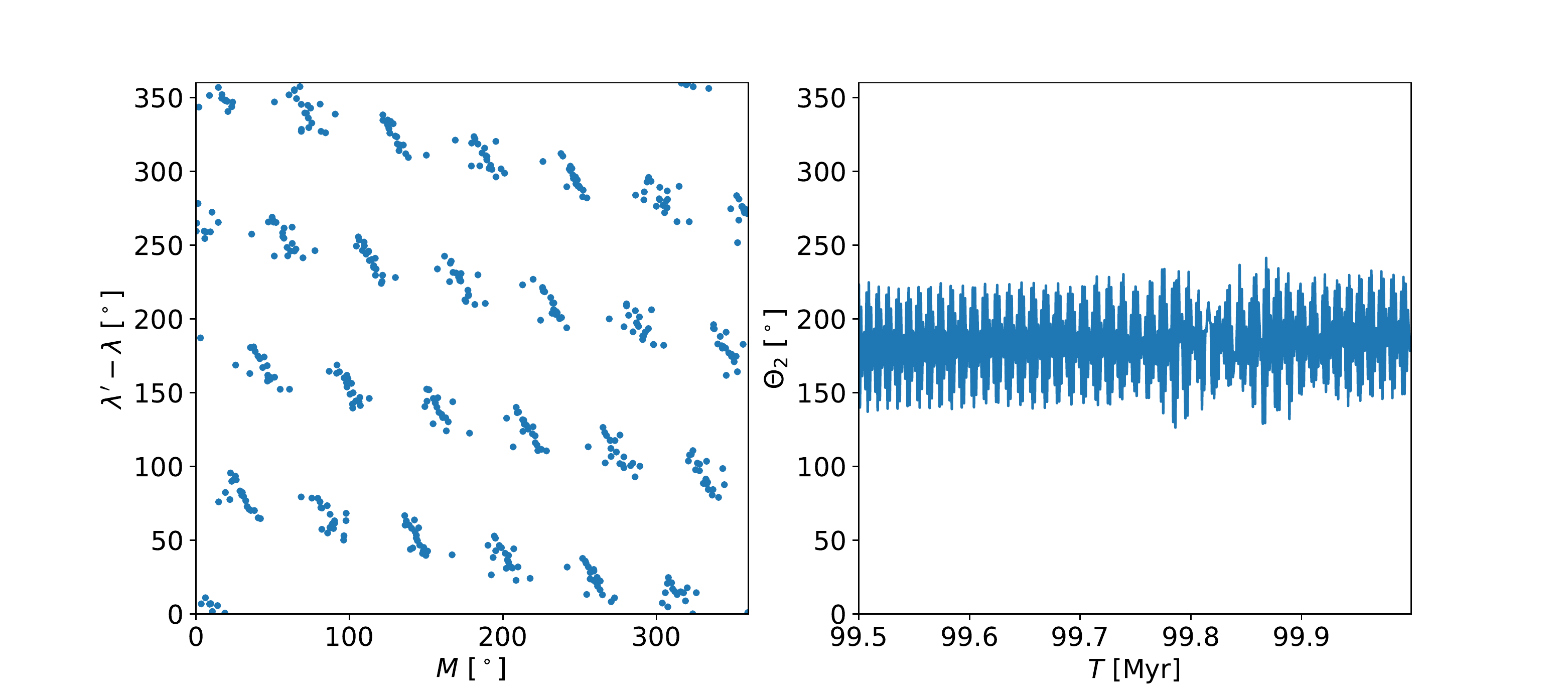}
    \caption{4:3 MMR between planet 3 and 4 in planetary system 192 (7PW model, $2.5\,\mathrm{M_\odot}$ host star) between $99.5$--$100\,\mathrm{Myr}$.}
    \label{fig:MMR_p_sys_192_7PW_2_5MSun}
\end{figure}

\begin{figure}
    \centering
    \includegraphics[width=0.5\textwidth]{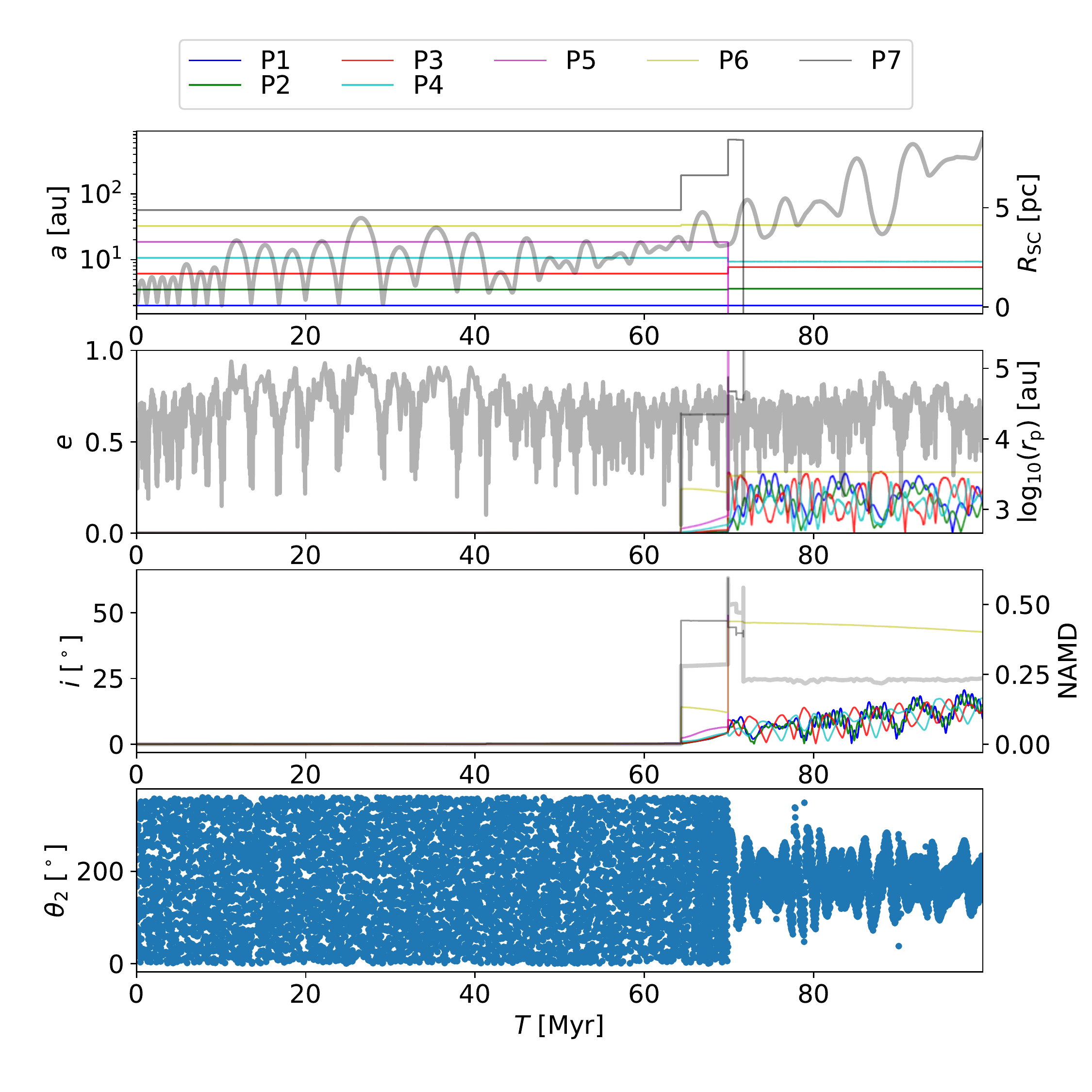}
    \caption{As Fig.~\ref{fig:elements_p_sys_122_7PC_1_5MSun} but for planetary system 192 (7PW model, $2.5\,\mathrm{M_\odot}$ host star). The resonance angle is shown for $p=3$, $q=1$ and planet pair 3/4.}
    \label{fig:elements_p_sys_192_7PW_2_5MSun}
\end{figure}

\newpage

\section{Additional Material}

\begin{table*}
    \centering
    \caption{Extract from the simulation results for the 3P planetary system model around $1.5\,\mathrm{M}_\odot$ host stars.  A particle ID equal to 0 corresponds to the system's central star. Ejected planets were omitted, as were systems where no planets remained.}
    \begin{tabular}{l|cccccccccccccc}
        \hline
System & Particle & a & e & i & x & y & z & vx & vy & vz & Particle & Stability \\ 
ID & ID & [au] & & [rad] & [au] & [au] & [au] & [au/d] & [au/d] & [au/d] & Mass [$\mathrm{M}_\odot$] & \\
\hline
0 & 0 & nan & nan & nan & -0.000 & -0.001 & -0.001 & -0.000 & 0.000 & -0.000 & 1.500E+00 & 0.91 \\
0 & 1 &  2.00 &  0.17 &  0.26 & -1.641 & -1.167 & 0.201 & 0.011 & -0.010 & 0.003 & 9.546E-06 & 0.91\\
0 & 2 &  6.12 &  0.17 &  0.53 & -5.786 & -2.263 & 3.397 & 0.004 & -0.006 & 0.001 & 9.546E-06 & 0.91\\
1 & 0 & nan & nan & nan & -0.186 & -0.052 & -0.267 & -0.000 & -0.000 & -0.000 & 1.500E+00 & 0.94 \\
1 & 1 &  2.00 &  0.00 &  0.00 & -1.667 & -1.396 & -0.267 & 0.010 & -0.011 & -0.000 & 9.546E-06 & 0.94\\
1 & 2 &  6.10 &  0.00 &  0.00 & 5.907 & 0.325 & -0.267 & -0.001 & 0.009 & -0.000 & 9.546E-06 & 0.94\\
1 & 3 & 18.63 &  0.00 &  0.00 & 1.978 & 18.454 & -0.266 & -0.005 & 0.001 & -0.000 & 9.546E-06 & 0.94\\
2 & 0 & nan & nan & nan & 0.005 & -0.003 & -0.005 & 0.000 & 0.000 & -0.000 & 1.500E+00 & 0.89 \\
2 & 1 &  2.00 &  0.03 &  0.06 & -2.023 & -0.287 & 0.011 & 0.002 & -0.014 & 0.001 & 9.546E-06 & 0.89\\
2 & 2 &  6.10 &  0.05 &  0.05 & -5.413 & -2.626 & 0.248 & 0.003 & -0.008 & 0.000 & 9.546E-06 & 0.89\\
2 & 3 & 17.09 &  0.29 &  0.06 & 16.191 & 13.060 & -0.809 & -0.003 & 0.003 & 0.000 & 9.546E-06 & 0.89\\
3 & 0 & nan & nan & nan & -0.001 & 0.001 & 0.001 & -0.000 & 0.000 & -0.000 & 1.500E+00 & 0.94 \\
3 & 1 &  2.00 &  0.01 &  0.04 & -1.089 & -1.686 & 0.059 & 0.013 & -0.008 & 0.000 & 9.546E-06 & 0.94\\
3 & 2 &  6.11 &  0.01 &  0.04 & -3.369 & 5.146 & -0.249 & -0.007 & -0.005 & 0.000 & 9.546E-06 & 0.94\\
3 & 3 & 19.11 &  0.05 &  0.02 & -18.023 & -8.944 & -0.187 & 0.002 & -0.004 & 0.000 & 9.546E-06 & 0.94\\
4 & 0 & nan & nan & nan & 0.057 & 0.476 & -0.537 & 0.000 & 0.000 & -0.000 & 1.500E+00 & 0.93 \\
4 & 1 &  2.00 &  0.08 &  0.28 & 0.501 & 2.328 & -0.990 & -0.015 & 0.002 & -0.002 & 9.546E-06 & 0.93\\
4 & 2 &  8.01 &  0.22 &  1.30 & -2.771 & 0.292 & 8.495 & 0.003 & -0.005 & 0.000 & 9.546E-06 & 0.93\\
5 & 0 & nan & nan & nan & -0.013 & -0.021 & 0.010 & -0.000 & -0.000 & 0.000 & 1.500E+00 & 0.93 \\
5 & 1 &  2.00 &  0.00 &  0.01 & 1.927 & -0.508 & 0.012 & 0.004 & 0.014 & 0.000 & 9.546E-06 & 0.93\\
5 & 2 &  6.10 &  0.00 &  0.01 & 5.687 & -2.214 & 0.005 & 0.003 & 0.008 & 0.000 & 9.546E-06 & 0.93\\
5 & 3 & 18.64 &  0.00 &  0.01 & -6.518 & -17.419 & -0.069 & 0.005 & -0.002 & 0.000 & 9.546E-06 & 0.93\\
6 & 0 & nan & nan & nan & 0.004 & 0.004 & -0.001 & 0.000 & -0.000 & -0.000 & 1.500E+00 & 0.94 \\
6 & 1 &  2.00 &  0.00 &  0.00 & 0.069 & 2.003 & -0.002 & -0.015 & 0.000 & 0.000 & 9.546E-06 & 0.94\\
6 & 2 &  6.10 &  0.00 &  0.00 & 5.518 & 2.622 & -0.004 & -0.004 & 0.008 & -0.000 & 9.546E-06 & 0.94\\
6 & 3 & 18.63 &  0.00 &  0.00 & 13.892 & -12.414 & 0.047 & 0.003 & 0.004 & -0.000 & 9.546E-06 & 0.94\\
7 & 0 & nan & nan & nan & 0.015 & -0.007 & 0.004 & 0.000 & -0.000 & 0.000 & 1.500E+00 & 0.93 \\
7 & 1 &  2.00 &  0.00 &  0.00 & -1.805 & 0.821 & 0.000 & -0.006 & -0.014 & 0.000 & 9.546E-06 & 0.93\\
7 & 2 &  6.10 &  0.00 &  0.00 & 4.142 & 4.493 & 0.010 & -0.006 & 0.006 & -0.000 & 9.546E-06 & 0.93\\
7 & 3 & 18.63 &  0.00 &  0.01 & 10.357 & -15.517 & 0.053 & 0.004 & 0.003 & 0.000 & 9.546E-06 & 0.93\\
8 & 0 & nan & nan & nan & 0.000 & 0.000 & 0.001 & -0.000 & 0.000 & 0.000 & 1.500E+00 & 0.90 \\
8 & 1 &  1.94 &  0.47 &  0.06 & -0.246 & -0.996 & 0.017 & 0.024 & -0.006 & 0.002 & 9.546E-06 & 0.90\\
8 & 2 &  6.96 &  0.06 &  0.17 & 2.961 & -5.973 & -0.340 & 0.007 & 0.003 & -0.001 & 9.546E-06 & 0.90\\
8 & 3 & 28.75 &  0.70 &  0.09 & -35.177 & -30.718 & 4.088 & 0.002 & -0.001 & -0.000 & 9.546E-06 & 0.90\\
9 & 0 & nan & nan & nan & 0.025 & 0.029 & -0.006 & 0.000 & 0.000 & -0.000 & 1.500E+00 & 0.92 \\
9 & 1 &  2.00 &  0.00 &  0.00 & 1.884 & 0.768 & -0.006 & -0.006 & 0.014 & 0.000 & 9.546E-06 & 0.92\\
9 & 2 &  6.10 &  0.00 &  0.00 & 2.725 & 5.504 & -0.006 & -0.008 & 0.004 & 0.000 & 9.546E-06 & 0.92\\
9 & 3 & 18.63 &  0.00 &  0.00 & -12.550 & 13.786 & 0.013 & -0.004 & -0.003 & 0.000 & 9.546E-06 & 0.92\\
10 & 0 & nan & nan & nan & 0.019 & 0.014 & 0.003 & 0.000 & 0.000 & 0.000 & 1.500E+00 & 0.93 \\
10 & 1 &  2.00 &  0.00 &  0.00 & -0.421 & -1.937 & 0.004 & 0.015 & -0.003 & -0.000 & 9.546E-06 & 0.93\\
10 & 2 &  6.10 &  0.00 &  0.00 & 4.214 & 4.448 & -0.000 & -0.006 & 0.006 & 0.000 & 9.546E-06 & 0.93\\
10 & 3 & 18.63 &  0.00 &  0.00 & 4.954 & 17.982 & 0.032 & -0.005 & 0.001 & 0.000 & 9.546E-06 & 0.93\\
11 & 0 & nan & nan & nan & 0.002 & 0.006 & -0.007 & -0.000 & 0.000 & -0.000 & 1.500E+00 & 0.92 \\
11 & 1 &  2.00 &  0.00 &  0.00 & -1.238 & -1.572 & -0.005 & 0.012 & -0.009 & 0.000 & 9.546E-06 & 0.92\\
11 & 2 &  6.11 &  0.00 &  0.01 & 5.810 & -1.883 & -0.049 & 0.003 & 0.008 & -0.000 & 9.546E-06 & 0.92\\
12 & 0 & nan & nan & nan & 0.000 & -0.000 & 0.000 & -0.000 & -0.000 & 0.000 & 1.500E+00 & 0.93 \\
12 & 1 &  2.00 &  0.00 &  0.00 & 0.136 & -1.995 & -0.001 & 0.015 & 0.001 & 0.000 & 9.546E-06 & 0.93\\
12 & 2 &  6.10 &  0.00 &  0.00 & 1.587 & -5.895 & -0.003 & 0.008 & 0.002 & 0.000 & 9.546E-06 & 0.93\\
12 & 3 & 18.63 &  0.00 &  0.00 & -1.078 & 18.600 & 0.029 & -0.005 & -0.000 & -0.000 & 9.546E-06 & 0.93\\
13 & 0 & nan & nan & nan & 0.003 & -0.001 & -0.004 & -0.000 & -0.000 & -0.000 & 1.500E+00 & 0.93 \\
13 & 1 &  2.00 &  0.00 &  0.00 & 1.722 & -1.023 & 0.001 & 0.008 & 0.013 & -0.000 & 9.546E-06 & 0.93\\
13 & 2 &  6.10 &  0.00 &  0.00 & -6.078 & 0.536 & -0.008 & -0.001 & -0.008 & 0.000 & 9.546E-06 & 0.93\\
13 & 3 & 18.63 &  0.00 &  0.00 & 6.921 & -17.302 & 0.008 & 0.005 & 0.002 & 0.000 & 9.546E-06 & 0.93\\
15 & 0 & nan & nan & nan & 0.004 & 0.042 & -0.133 & -0.000 & 0.000 & -0.000 & 1.500E+00 & 0.92 \\
15 & 1 &  2.00 &  0.00 &  0.00 & 1.974 & 0.384 & -0.133 & -0.003 & 0.015 & 0.000 & 9.546E-06 & 0.92\\
15 & 2 &  6.10 &  0.00 &  0.00 & 5.116 & -3.294 & -0.133 & 0.005 & 0.007 & 0.000 & 9.546E-06 & 0.92\\
15 & 3 & 18.64 &  0.00 &  0.00 & -16.439 & -8.721 & -0.170 & 0.002 & -0.004 & -0.000 & 9.546E-06 & 0.92\\
\vdots & \vdots & \vdots & \vdots & \vdots & \vdots & \vdots & \vdots & \vdots & \vdots & \vdots & \vdots & \vdots \\
        \hline
    \end{tabular}
    \label{tab:results_1_5MSun_3P}
\end{table*}

\begin{table*}
    \centering
    \caption{Extract from the simulation results for the 7PW planetary system model around $2.5\,\mathrm{M}_\odot$ host stars.  A particle ID equal to 0 corresponds to the system's central star. Ejected planets were omitted, as were systems where no planets remained.}
    \begin{tabular}{l|cccccccccccccc}
        \hline
System & Particle & a & e & i & x & y & z & vx & vy & vz & Particle & Stability \\ 
ID & ID & [au] & & [rad] & [au] & [au] & [au] & [au/d] & [au/d] & [au/d] & Mass [$\mathrm{M}_\odot$] & \\
\hline
0 & 0 & nan & nan & nan & -0.001 & -0.001 & 0.001 & -0.000 & -0.000 & 0.000 & 2.500E+00 & 0.16 \\ 
0 & 1 &  1.23 &  0.36 &  0.53 & -1.112 & -0.953 & 0.062 & 0.015 & -0.009 & -0.010 & 9.546E-06 & 0.16\\ 
0 & 2 &  9.36 &  0.16 &  0.51 & -4.971 & 8.082 & 5.173 & -0.006 & -0.005 & 0.001 & 9.546E-06 & 0.16\\ 
0 & 3 &  3.20 &  0.19 &  0.12 & 3.415 & 0.233 & -0.273 & -0.003 & 0.014 & 0.001 & 9.546E-06 & 0.16\\ 
0 & 4 &  7.12 &  0.72 &  0.43 & 8.489 & -1.473 & 1.580 & 0.006 & 0.004 & 0.003 & 9.546E-06 & 0.16\\ 
0 & 5 & 26.48 &  0.59 &  0.38 & -16.714 & 28.949 & 4.355 & -0.002 & -0.003 & -0.001 & 9.546E-06 & 0.16\\ 
0 & 6 & 37.90 &  0.20 &  0.51 & 38.351 & 22.718 & -3.512 & -0.002 & 0.003 & 0.002 & 9.546E-06 & 0.16\\ 
0 & 7 & 139.55 &  0.62 &  0.16 & 223.221 & -2.257 & -32.354 & -0.000 & 0.001 & 0.000 & 9.546E-06 & 0.16\\ 
1 & 0 & nan & nan & nan & -0.075 & -0.039 & 0.036 & -0.000 & -0.000 & 0.000 & 2.500E+00 & 0.90 \\ 
1 & 1 &  2.00 &  0.02 &  0.09 & 0.111 & 1.989 & 0.188 & -0.019 & 0.002 & -0.001 & 9.546E-06 & 0.90\\ 
1 & 2 &  3.49 &  0.03 &  0.09 & -2.978 & -1.795 & -0.268 & 0.008 & -0.013 & -0.000 & 9.546E-06 & 0.90\\ 
1 & 3 &  6.10 &  0.05 &  0.09 & 0.129 & -6.055 & -0.339 & 0.011 & -0.000 & 0.001 & 9.546E-06 & 0.90\\ 
1 & 4 & 10.70 &  0.12 &  0.09 & -9.250 & -2.138 & -0.685 & 0.002 & -0.009 & -0.000 & 9.546E-06 & 0.90\\ 
1 & 5 & 18.54 &  0.18 &  0.08 & 8.254 & 14.019 & 1.340 & -0.007 & 0.003 & -0.000 & 9.546E-06 & 0.90\\ 
1 & 6 & 34.84 &  0.13 &  0.07 & 7.110 & -30.189 & 0.744 & 0.005 & 0.002 & 0.000 & 9.546E-06 & 0.90\\ 
2 & 0 & nan & nan & nan & -0.000 & 0.000 & 0.000 & -0.000 & -0.000 & 0.000 & 2.500E+00 & 0.93 \\ 
2 & 1 &  2.00 &  0.00 &  0.06 & -0.365 & -1.964 & 0.098 & 0.019 & -0.003 & 0.001 & 9.546E-06 & 0.93\\ 
2 & 2 &  3.49 &  0.00 &  0.06 & -0.493 & 3.453 & -0.206 & -0.014 & -0.002 & -0.000 & 9.546E-06 & 0.93\\ 
2 & 3 &  6.10 &  0.00 &  0.06 & 5.546 & -2.536 & 0.306 & 0.005 & 0.010 & -0.000 & 9.546E-06 & 0.93\\ 
2 & 4 & 10.67 &  0.00 &  0.06 & 10.500 & 1.926 & 0.188 & -0.001 & 0.008 & -0.001 & 9.546E-06 & 0.93\\ 
2 & 5 & 18.63 &  0.00 &  0.06 & -18.587 & 0.176 & -0.459 & -0.000 & -0.006 & 0.000 & 9.546E-06 & 0.93\\ 
2 & 6 & 32.63 &  0.00 &  0.06 & -26.068 & -19.737 & 0.814 & 0.003 & -0.004 & 0.000 & 9.546E-06 & 0.93\\ 
2 & 7 & 61.95 &  0.12 &  0.05 & -14.436 & 53.361 & 1.607 & -0.004 & -0.001 & 0.000 & 9.546E-06 & 0.93\\ 
5 & 0 & nan & nan & nan & -0.012 & -0.006 & 0.005 & -0.000 & -0.000 & 0.000 & 2.500E+00 & 0.89 \\ 
5 & 1 &  2.00 &  0.00 &  0.04 & 1.984 & -0.003 & -0.058 & 0.000 & 0.019 & 0.000 & 9.546E-06 & 0.89\\ 
5 & 2 &  3.49 &  0.00 &  0.04 & -3.199 & -1.437 & 0.072 & 0.006 & -0.013 & -0.001 & 9.546E-06 & 0.89\\ 
5 & 3 &  6.10 &  0.00 &  0.05 & 2.113 & 5.739 & 0.114 & -0.010 & 0.004 & 0.000 & 9.546E-06 & 0.89\\ 
5 & 4 & 10.70 &  0.01 &  0.07 & 10.740 & 0.963 & -0.348 & -0.001 & 0.008 & 0.000 & 9.546E-06 & 0.89\\ 
5 & 5 & 18.87 &  0.06 &  0.10 & -13.367 & -13.405 & -0.398 & 0.004 & -0.005 & -0.001 & 9.546E-06 & 0.89\\ 
5 & 6 & 32.69 &  0.08 &  0.08 & -15.408 & 26.436 & 2.387 & -0.005 & -0.002 & -0.000 & 9.546E-06 & 0.89\\ 
5 & 7 & 50.95 &  0.28 &  0.07 & -37.901 & -3.118 & 1.144 & -0.000 & -0.005 & -0.000 & 9.546E-06 & 0.89\\ 
6 & 0 & nan & nan & nan & -0.002 & 0.002 & 0.002 & -0.000 & -0.000 & -0.000 & 2.500E+00 & 0.00 \\ 
6 & 1 &  2.05 &  0.56 &  0.97 & 1.795 & 0.736 & -0.952 & 0.009 & 0.014 & 0.006 & 9.546E-06 & 0.00\\ 
6 & 3 &  6.06 &  0.37 &  0.72 & 5.220 & 6.275 & 0.430 & -0.004 & 0.004 & 0.005 & 9.546E-06 & 0.00\\ 
6 & 4 & 23.56 &  0.69 &  1.06 & 0.258 & -34.796 & -18.666 & 0.001 & 0.001 & -0.002 & 9.546E-06 & 0.00\\ 
6 & 5 & 17.11 &  0.15 &  0.10 & 18.423 & -2.285 & -1.782 & 0.001 & 0.006 & -0.000 & 9.546E-06 & 0.00\\ 
6 & 6 & 26.71 &  0.17 &  0.30 & -6.645 & 20.310 & 5.730 & -0.006 & -0.002 & -0.001 & 9.546E-06 & 0.00\\ 
6 & 7 & 133.92 &  0.36 &  1.09 & 104.096 & 66.677 & -36.169 & -0.001 & 0.001 & 0.002 & 9.546E-06 & 0.00\\
7 & 0 & nan & nan & nan & 0.002 & -0.001 & -0.002 & -0.000 & 0.000 & 0.000 & 2.500E+00 & 0.62 \\ 
7 & 1 &  1.85 &  0.49 &  0.69 & -1.359 & -2.317 & -0.208 & 0.007 & -0.006 & -0.008 & 9.546E-06 & 0.62\\ 
9 & 0 & nan & nan & nan & 0.000 & -0.000 & 0.001 & 0.000 & -0.000 & 0.000 & 2.500E+00 & 0.31 \\ 
9 & 2 &  3.30 &  0.57 &  0.24 & 3.111 & 2.797 & -0.723 & -0.002 & 0.011 & -0.003 & 9.546E-06 & 0.31\\ 
9 & 5 & 90.06 &  0.95 &  1.15 & -33.978 & 153.282 & -75.976 & -0.000 & 0.000 & -0.000 & 9.546E-06 & 0.31\\ 
9 & 6 & 100.61 &  0.60 &  0.21 & -47.834 & -145.928 & 10.892 & 0.001 & 0.000 & 0.000 & 9.546E-06 & 0.31\\ 
10 & 0 & nan & nan & nan & -0.000 & 0.000 & 0.000 & 0.000 & 0.000 & 0.000 & 2.500E+00 & 0.91 \\ 
10 & 1 &  2.00 &  0.00 &  0.00 & 0.492 & 1.939 & -0.001 & -0.019 & 0.005 & -0.000 & 9.546E-06 & 0.91\\ 
10 & 2 &  3.49 &  0.00 &  0.00 & -1.407 & 3.198 & -0.009 & -0.013 & -0.006 & -0.000 & 9.546E-06 & 0.91\\ 
10 & 3 &  6.10 &  0.00 &  0.00 & 2.755 & 5.447 & 0.005 & -0.010 & 0.005 & -0.000 & 9.546E-06 & 0.91\\ 
10 & 4 & 10.67 &  0.00 &  0.00 & -9.761 & -4.294 & -0.037 & 0.003 & -0.008 & 0.000 & 9.546E-06 & 0.91\\ 
10 & 5 & 18.64 &  0.00 &  0.00 & -18.001 & 4.714 & -0.070 & -0.002 & -0.006 & -0.000 & 9.546E-06 & 0.91\\ 
10 & 6 & 32.52 &  0.00 &  0.01 & 32.360 & -3.619 & 0.148 & 0.001 & 0.005 & 0.000 & 9.546E-06 & 0.91\\ 
10 & 7 & 56.91 &  0.01 &  0.01 & 27.626 & -49.813 & -0.455 & 0.003 & 0.002 & 0.000 & 9.546E-06 & 0.91\\ 

\vdots & \vdots & \vdots & \vdots & \vdots & \vdots & \vdots & \vdots & \vdots & \vdots & \vdots & \vdots & \vdots \\

        \hline
    \end{tabular}
    \label{tab:results_2_5MSun_7PW}
\end{table*}

\bsp	
\label{lastpage}
\end{document}